\documentclass[twocolumn]{revtex4}
\usepackage[]{graphicx} 
\usepackage{amsmath}
\usepackage{multirow}
\begin{document}
\title{Non-hyperuniform metastable states around a disordered hyperuniform state of densely packed spheres: stochastic density functional theory at strong coupling}

\author{Hiroshi Frusawa}
\email{frusawa.hiroshi@kochi-tech.ac.jp}
\affiliation{Laboratory of Statistical Physics, Kochi University of Technology, Tosa-Yamada, Kochi 782-8502, Japan.}

\date{\today}
\begin{abstract}
Disordered and hyperuniform structures of densely packed spheres near and at jamming are characterized by vanishing of long-wavelength density fluctuations, or equivalently by long-range power-law decay of the direct correlation function (DCF).
We focus on previous simulation results that exhibit degradation of hyperuniformity in jammed structures while maintaining the long-range nature of the DCF to a certain length scale.
Here we demonstrate that a field-theoretic formulation of the stochastic density functional theory is relevant to explore the degradation mechanism.
The strong-coupling expansion method of the stochastic density functional theory is developed to obtain the metastable chemical potential considering intermittent fluctuations in dense packings.
The metastable chemical potential yields an analytical form of the metastable DCF that has a short-range cutoff inside the sphere while retaining the long-range power-law behavior.
It is confirmed that the metastable DCF provides zero-wavevector limit of structure factor in quantitative agreement with the previous simulation results of degraded hyperuniformity.
We can also predict the emergence of soft modes localized at the particle scale from plugging this metastable DCF into the linearized Dean-Kawasaki equation, a stochastic density functional equation.
\end{abstract}


\maketitle
\section{Introduction}
Hyperuniformity is characterized by density fluctuations that decrease to zero at the longest scales \cite{t revjcp,t physrep,t mat,h mat2020,t first,t donev1,t donev2,t disk,t void1,t void2,t rattler,t marginal,t dcf,t slowing,t kim,t real,t preprint}.
We have observed the hyperuniform states in a variety of complex soft matter systems, including foams, polymer blends, colloidal suspensions and biological tissues (see \cite{t revjcp,t physrep} for reviews).
It has also been found that non-crystalline materials with hyperuniformity have unique physical properties such as high-density transparency and isotropic filtration of elastic or electromagnetic waves \cite{t revjcp,t physrep,t mat,h mat2020}.
Consequently, considerable attention has been given to the disordered hyperuniform materials fabricated at the micro-and nano-scales, because of their potential importance for applications in photonics, electronics, and structural components with novel properties \cite{t mat,h mat2020}.

Here we focus on computer glasses among the disordered hyperuniform systems.
Recent methodological developments allow us to create computer glasses in an experimentally relevant regime \cite{t revjcp,t physrep,t mat,h mat2020,t first,t donev1,t donev2,t disk,t void1,t void2,t rattler,t marginal,t dcf,t slowing,t kim,t real,t preprint,hecke rev,liu rev,wyart rev,lub2017,lub2018,ikeda2017,ikeda2018,ikeda2020 prr,ikeda2020 jcp,ikeda2020 sm,lerner2013,lerner2016,lerner2020 prl,lerner2020 pre,lerner2020 pnas,urbani2021,berthier2016,szamel nc,berthier2019,tan sm2021,ozawa,chieco,silbert,olsson,ikeda non1,ikeda non2}, and yet the disordered hyperuniformity at jamming has not always been realized \cite{t revjcp,t physrep,ozawa,chieco,silbert,olsson,ikeda non1,ikeda non2}.
The emergence of hyperuniformity depends on the preparation protocols, partly because of a significantly long computational time that is required to determine the configurations near and at jamming \cite{t revjcp,t physrep,t first,t donev1,t donev2,t disk,t void1,t void2,t rattler,t marginal,t dcf,t slowing}.

On the one hand, some simulation studies have demonstrated the hyperuniformity in densely packed spheres: the structure factor $S(k)$ in a hyperuniform state exhibits a non-trivial linear dependence on the wavevector magnitude $k$ in the low-wavevector range near and at jamming (i.e., $S(k)\sim k\quad(k\geq 0)$), and the zero-wavevector limit of the structure factor $S(0)$ eventually vanishes at jamming \cite{t revjcp,t physrep,t first,t donev1,t donev2,t disk,t void1,t void2,t rattler,t marginal,t dcf,t slowing}.
These results indicate not only the existence of long-range order, but also the complete suppression of density fluctuations over the system scale.

Meanwhile, other simulation studies near and at jamming \cite{ozawa,chieco,silbert,olsson,ikeda non1,ikeda non2} provide non-vanishing structure factor at zero wavevector.
The degradation of hyperuniformity is that either a saturation or an upturn is observed for $S(k)$ at the lowest values of $k$, despite the linear relation above a crossover wavevector $k_c$ \cite{ozawa,chieco,silbert,olsson,ikeda non1,ikeda non2}:
\begin{flalign}
&S(k)\sim k\quad(k\geq k_c),\nonumber\\
&\lim_{k\rightarrow 0}S(k)\geq S(k_c).
\label{non hyper structure}
\end{flalign}
It has also been demonstrated that $S(0)$ is weakly dependent on the density variation \cite{ozawa,chieco,silbert,olsson,ikeda non1,ikeda non2}.

The quantitative difference between the non-hyperuniform and hyperuniform states can be seen from the inverse of the zero-wavevector structure factors.
While the non-vanishing values of $S(0)$ {\itshape due to the incomplete linear-dependence of $S(k)$} are in the range of \cite{ozawa,chieco,silbert,olsson,ikeda non1,ikeda non2}
\begin{equation}
10^{2}<\frac{1}{S(0)}<10^{3},
\label{nonhyper result}
\end{equation}
the hyperuniform computer glasses have been characterized by the inequality,
\begin{equation}
\frac{1}{S(0)}>10^{4},
\label{hyper result}
\end{equation}
irrespective of system details \cite{t revjcp,t physrep,t first,t donev1,t donev2,t disk,t void1,t void2,t rattler,t marginal,t dcf,t slowing}.

In terms of density-density correlation functions in real space, hyperuniformity is a kind of inverted critical phenomenon.
It is among critical phenomena in normal fluids that total correlation functions are long-ranged at critical points, accompanied by the diverging behaviors of density fluctuation and isothermal compressibility, while keeping the direct correlation function (DCF) short-ranged.
In contrast, the inverted critical phenomenon is that the hyperuniform DCF is long-ranged in correspondence with the vanishing isothermal compressibility, despite the absence of long-range behavior for the total correlation function \cite{t revjcp,t physrep,t first,t donev1,t donev2,t disk,t void1,t void2,t rattler,t marginal,t dcf,t slowing, t preprint}.

More concretely, the long-range behavior of the hyperuniform DCF $c({\bf r})$ is described by the power law as follows \cite{t revjcp,t physrep,t first,t donev1,t donev2,t dcf, t preprint}:
\begin{equation}
c({\bf r})\sim \frac{1}{r^2}\quad(r>\sigma),
\label{hyper dcf}
\end{equation}
where $|{\bf r}|=r$ and $\sigma$ denotes the sphere diameter. The long-range decay of $c({\bf r})$ reads $c(k)\sim k^{-1}$ in the Fourier space, which is equivalent to the the linear behavior $S(k)\sim k$ due to the following relation between $S(k)$ and $c(k)$:
\begin{flalign}
\frac{1}{S(k)}=1-nc(k),
\label{fourier dcf}
\end{flalign}
with $n$ being the spatially averaged density of spheres. Furthermore, in disordered packings of hard spheres, the hyperuniform DCF for $r\leq\sigma$ satisfies another power law \cite{t dcf},
\begin{equation}
c({\bf r})\sim \frac{1}{r}\quad(r\leq\sigma),
\label{small hyper dcf}
\end{equation}
which is divergent at small $r$.

It follows from eqs. (\ref{non hyper structure}) and (\ref{fourier dcf}) that the non-hyperuniform DCF $c({\bf r})$ at jamming satisfies the scaling relation (\ref{hyper dcf}) over a finite range.
In other words, the violation of hyperuniformity occurs while maintaining the long-range behavior to a length scale $L_c$: eq. (\ref{hyper dcf}) holds in the range of $\sigma <r\leq L_c$ \cite{ozawa,chieco,silbert,olsson,ikeda non1,ikeda non2} where
\begin{flalign}
6\leq\frac{L_c}{\sigma}\leq 10.
\label{lc result}
\end{flalign}
It is to be noted that simulation results of hyperuniform systems are also likely to provide the finiteness represented by eq. (\ref{lc result}) \cite{t revjcp,t physrep,t first,t donev1,t donev2,t disk,t void1,t void2,t rattler,t marginal,t dcf,t slowing}, for we have computational difficulty obtaining the scaling behavior ($c({\bf r})\sim r^{-2}$) over $L_c$ from the structure factor, irrespective of whether the computer glass is in a hyperuniform or non-hyperuniform state.

This common feature of the long-range behavior (eq. (\ref{hyper dcf})) in the hyperuniform and non-hyperuniform DCFs raises the question of what causes the difference between eqs. (\ref{nonhyper result}) and (\ref{hyper result}). Accordingly, it is the purpose of this paper to reveal the underlying mechanism behind the difference between emergence and degradation of hyperuniformity.
To this end, we formulate an analytical form of the non-vanishing zero-wavevector structure factor that satisfies eq. (\ref{nonhyper result}) under the condition of eq. (\ref{lc result}).
A key ingredient of our formulation is {\itshape the strong-coupling approximation of the stochastic density functional theory (DFT) \cite{witt,dean,seifert,kim,jac,das2015,podgornik,demery,kruger,lut science,frusawa1,frusawa2,goles} which can consider intermittent fluctuations while fixing the density field at a given distribution of dense packings near and at jamming}.

The remainder of this paper consists of two parts.
In the former part of Secs. II to IV, the problem to be addressed is defined.
Section II provides a theoretical background as to why the stochastic DFT should be brought into the problem on the fluctuation-induced non-hyperuniformity.
In Sec. III, the basic formulation of the stochastic DFT is presented, focusing on the definition of metastable states.
Then, we find that the stochastic DFT allows us to relate the metastable zero-wavevector structure factor $S^*(0)$ to a potential energy $\lambda^*$ per particle, which will be referred to as metastable chemical potential.
In Sec. IV, the non-hyperuniform state on target is specified using Table 1, a classification list of hyperuniform and non-hyperuniform systems.

We see from the system specification that the non-hyperuniformity of our concern requires the short-range cutoff of the metastable DCF $c^*({\bf r})$, as well as a drop in the long-ranged DCF for $r>L_c$.
Our primary goal is to derive the short-range cutoff of the DCF by developing the strong-coupling approximation for the stochastic DFT.

The latter part of this paper presents the results and discussion regarding the metastable DCF $c^*({\bf r})$.
Before entering the main results, Sec. V compares the stochastic DFT with the equilibrium DFT \cite{ry,evans,singh,likos,lut1,singh1985,baus,woly1987,dasgupta1992,bagchi,das2001,munakata, sood,dasgupta2008,das2012,das2016,odagaki,das2020}
in terms of $S^*(0)$.
It is demonstrated as a preliminary result that the resulting forms of $S^*(0)$ in the equilibrium and stochastic DFTs coincide with each other as far as the Gaussian approximation of the stochastic DFT is adopted.
Section VI provides the metastable DCF expressed by the Mayer-type function form, hence verifying the cutoff for the metastable DCF $c^*({\bf r})$ inside the sphere.
As a consequence, we confirm that $S^*(0)$ satisfies the relation (\ref{nonhyper result}), instead of eq.
(\ref{hyper result}).
In Sec. VII, the coupling constant $\gamma$ to represent the strength of interactions is introduced using the hyperuniform DCF $c({\bf r})$, and it is shown that the $1/\gamma$ expansion method becomes equivalent to the virial-type one at strong coupling of $\gamma\gg 1$.
Correspondingly, the interaction term in the metastable chemical potential $\lambda^*$ is expressed by the above metastable DCF $c^*({\bf r})$.
In Sec. VIII, the stochastic density functional equation clarifies that the short-range cutoff of the metastable DCF $c^*({\bf r})$ leads to the emergence of dynamic softening at the particle scale: the interaction-induced restoring force against density fluctuations around a metastable state vanishes within the scale of spherical diameter $\sigma$.
The microscopic mechanism behind the soft modes is also discussed in connection with previous simulation results.
Furthermore, both Fig. 3 and Table 2 summarize the static and dynamic results for comparing the equilibrium DFT, the stochastic DFT in the Gaussian approximation, and the stochastic DFT in the strong-coupling approximation.
Final remarks are given in Sec. IX.

\section{A theoretical background of the stochastic DFT}
This section is intended to provide a brief overview of theoretical approaches to jammed structures for explaining the relevance of the stochastic DFT  \cite{witt,dean,seifert,kim,jac,das2015,podgornik,demery,kruger,lut science,frusawa1,frusawa2,goles} to the degradation of hyperuniformity.

\subsection{Marginal stability and the free energy landscape}
There have been two conceptual approaches to address various issues on computer glasses, including the structure factors that vary depending on the protocols used \cite{t revjcp,t physrep,t first,t donev1,t donev2,t disk,t void1,t void2,t rattler,t marginal,t dcf,t slowing, t real}. One is the ensemble approach to investigate physically relevant packings based on a packing protocol selected. The other method is the geometric-structure approach for quantitative characterization of single-packing configurations to enumerate and classify the jammed structures.

On the one hand, the ensemble approach has involved the problem that the protocol-dependency of the occurrence frequency of jammed configurations leads to the ambiguity of weighing jammed states \cite{t revjcp,t physrep, t first,t donev1,t donev2,t disk,t void1,t void2,t rattler,t marginal,t dcf,t slowing, t kim,t real,t preprint}. Recently, however, the protocol-dependency problem is theoretically tackled: the canonical ensemble method is developed for a large number of allowed configurations to resolve the configuration realizability issue \cite{t real}.

The geometric-structure study, on the other hand, has distinguished three types of jamming for densely packed spheres \cite{t revjcp,t physrep, t rattler}: local, collective and strict jammings. These types of jamming are hierarchical in that local and collective jammings are prerequisites for collective and strict ones, respectively, as follows: (i) in locally jammed states, a particle cannot translate when the positions of all other spheres in the packing are fixed; (ii) in collectively jammed states, particles prevented from translating are further stable to uniform compression; (iii) strictly jammed packings are stable against both uniform and shear deformations.

The geometric-structure studies on various computer glasses have verified that the hyperuniformity emerges in either strictly or collectively jammed systems having isostaticity \cite{t revjcp,t physrep, t disk}. Here the isostatic configuration provides a mean contact number $2d$ per particle with $d$ being the spatial dimension, thereby enhancing the mechanical stability. To be noted, however, mechanical rigidity of jammed packings is a necessary but not sufficient condition for hyperuniformity.

It has been conjectured that any strictly jammed saturated infinite packing of identical spheres is hyperuniform; the conjecture excludes the existence of rattlers, or particles that are free to move in a confining cage, by definition of strictly jammed packings \cite{t revjcp,t physrep,t first,t marginal,t preprint}.
Conversely, it depends on simulation methods and conditions whether dense packings other than the strictly jammed ones, including the isostatic and collectively jammed systems, are hyperuniform or not.
The isostatic systems can be destabilized by cutting one particle contact unless disordered packings are strictly jammed. In other words, the isostaticity is a critical factor in the mechanical marginal stability \cite{t revjcp,t physrep, t marginal,hecke rev,liu rev,wyart rev,lub2017,lub2018,ikeda2017,ikeda2018,ikeda2020 prr,ikeda2020 jcp,ikeda2020 sm,lerner2013,lerner2016,lerner2020 prl,lerner2020 pre,lerner2020 pnas,urbani2021,berthier2016,szamel nc,berthier2019,tan sm2021, behringer rev, makse, b char2015,b franz2015,b char2016,b hexner,b franz2020,b char2020,b liu}.

Recent simulation results have demonstrated that thermal fluctuations in the marginal states are accompanied by intermittent rearrangements of particles \cite{hecke rev,liu rev,wyart rev,lub2017,lub2018,ikeda2017,ikeda2018,ikeda2020 prr,ikeda2020 jcp,ikeda2020 sm,lerner2013,lerner2016,lerner2020 prl,lerner2020 pre,lerner2020 pnas,urbani2021,berthier2016,szamel nc,berthier2019,tan sm2021}. As a consequence, the marginal systems become responsive to have low-frequency soft modes that are nonphononic and anharmonic.
For instance, quasi-localized modes coupled to an elastic matrix create soft spots composed of tens to hundreds of particles undergoing displacements \cite{hecke rev,liu rev,wyart rev,lub2017,lub2018,ikeda2017,ikeda2018,ikeda2020 prr,ikeda2020 jcp,ikeda2020 sm,lerner2013,lerner2016,lerner2020 prl,lerner2020 pre,lerner2020 pnas,urbani2021,berthier2016,szamel nc,berthier2019,tan sm2021}. The low-frequency soft modes exhibit similar behaviors, and the common features of marginal states have been related to the emergence of many local minima in the free-energy landscape \cite{ikeda2020 jcp,zamponi2012,zamponi2014,berthier prx,parisi2020}.

The similarity in anharmonic vibrations suggests that the ensemble of configurations visited by the slow dynamics could reveal the characteristics of marginal stability associated with the free-energy landscape, even though possible configurations depend on a protocol adopted \cite{t revjcp,t physrep,t first,t donev1,t donev2,t disk,t void1,t void2,t rattler,t marginal,t dcf,t slowing, t real}.

\subsection{The free-energy density functional: comparison between the stochastic and equilibrium DFTs}
For assessing the applicability of density functional approaches to the free-energy landscape in glassy systems, let us compare the stochastic DFT \cite{witt,dean,seifert,kim,jac,das2015,podgornik,demery,kruger,lut science,frusawa1,frusawa2,goles} and the equilibrium DFT, or the classical DFT conventionally used \cite{ry,evans,singh,likos,lut1,singh1985,baus,woly1987,dasgupta1992,bagchi,das2001,munakata, sood,dasgupta2008,das2012,das2016,odagaki,das2020}.

The equilibrium DFT \cite{ry,evans,singh,likos,lut1,singh1985,baus,woly1987,dasgupta1992,bagchi,das2001,munakata, sood,dasgupta2008,das2012,das2016,odagaki,das2020}, one of ensemble approaches, has been found relevant to investigate the free-energy landscape \cite{zamponi2012, zamponi2014,berthier prx, parisi2020, still jcp, still nat, heuer, ediger rev}. It has been demonstrated that metastable minima determined by the equilibrium DFT are not only correlated with the appearance of two-step relaxation and divergence of relaxation time, but are also directly connected with dynamical heterogeneity \cite{singh1985,baus,woly1987,dasgupta1992,bagchi,das2001,munakata, sood,dasgupta2008,das2012,das2016,odagaki,das2020}. In the equilibrium DFT, the metastable density profile $\rho^*({\bf r})$ has been approximated by a superposition of narrow Gaussian density profiles centered around a set of points forming an aperiodic lattice. The equilibrium DFT has properly identified the metastable state of the liquid having an inhomogeneous and aperiodic density as a local minimum of the equilibrium free-energy functional with respect to variation of the width parameter for the above Gaussian distribution \cite{singh1985,baus,woly1987,dasgupta1992,bagchi,das2001,munakata, sood,dasgupta2008,das2012,das2016,odagaki,das2020}.

However, the violation of perfect hyperuniformity has been beyond the scope of the equilibrium DFT.
Recently, the following three scenarios of imperfections have been proposed for demonstrating the degradation of hyperuniformity both theoretically and numerically \cite{t kim}: (i) uncorrelated point defects, (ii) stochastic particle displacements that are spatially correlated, and (iii) thermal excitations.
In this paper, we focus on the second scenario (ii) that is related to intermittent particle rearrangements in a contact network, a set of bonds connecting particles which are in contact with each other (see \cite{hecke rev,liu rev,wyart rev,lub2017,lub2018,behringer rev,makse} for reviews). The elastic nature of the contact network could be responsible for the above second scenario of non-hyperuniformity, or the spatially correlated displacements occurring stochastically; this will be discussed in Secs. VIII and IX, based on the results obtained herein.

From the stochastic DFT \cite{witt,dean,seifert,kim,jac,das2015,podgornik,demery,kruger,lut science,frusawa1,frusawa2,goles}, on the other hand, it is expected that the above second scenario (i.e., (ii) stochastic and spatially correlated displacements) could be described in terms of stochastic density dynamics. To see this, a brief review of the stochastic DFT is given below.

The stochastic DFT has been used as one of the most powerful tools for describing slowly fluctuating and/or intermittent phenomena, such as glassy dynamics, nucleation or pattern formation of colloidal particles, dielectric relaxation of Brownian dipoles, and even tumor growth (see \cite{witt} for a thorough review). The stochastic density functional equation, which has often been referred to as the Dean-Kawasaki equation \cite{witt,dean}, forms the basis of the stochastic DFT. It has been shown in various systems that the Dean-Kawasaki equation successfully describes the stochastic evolution of the instantaneous microscopic density field of overdamped Brownian particles. Of great practical use is the Dean-Kawasaki equation linearized with respect to density fluctuations around various reference density distributions \cite{witt,podgornik,demery,kruger,frusawa1,frusawa2,goles}.

As seen below, the stochastic DFT is formulated on the hybrid framework that combines the equilibrium DFT and the statistical field theory \cite{frusawa1,frusawa2,frusawa3}.
The hybrid framework allows us to investigate metastable states considering fluctuations as clarified below. In Sec. VIII, the stochastic DFT will also shed light on the dynamical properties of non-hyperuniformity.

\section{Basic formulation: the stochastic DFT}
This section shows that the stochastic DFT \cite{witt,dean,seifert,kim,jac,das2015,podgornik,demery,kruger,lut science,frusawa1,frusawa2,goles} is available to investigate the zero-wavevector limit of the structure factor $S^*(0)$ in a metastable state. It is not merely a review of the previous formulations \cite{frusawa1}, but rather a revisit for making it clear that the stochastic DFT serves as a systematic evaluation of $S^*(0)$: as demonstrated in Secs. VI and VII, we can evaluate the extent to which $S^*(0)$ is altered to the stochastic fluctuations around a hyperuniform state in a systematic manner.

First, the constrained free-energy functional $\mathcal{A}[\rho]$ is represented by the hybrid form using the functional and configurational integrals (Sec. IIIA). Next, we introduce the non-equilibrium excess chemical potential appearing in the stochastic DFT equation (Sec. IIIB). Third, the metastable state is defined based on the stochastic DFT (Sec. IIIC). Last, the metastable DCF $c^*({\bf r})$ is generated from the metastable chemical potential $\lambda^*$, thereby yielding $S^*(0)$ expressed by $c^*({\bf r})$ (Sec. IIID).

\subsection{Constrained free-energy functional $\mathcal{A}[\rho]$ in connection with the Fokker-Planck equation}
Let $\widehat{\rho}_N({\bf r},t)=\sum_{i=1}^N\delta[{\bf r}-{\bf r}_i(t)]$ be an instantaneous microscopic density of $N$-particle system where the position ${\bf r}_i(t)$ at time $t$ represents an instantaneous location of the $i$-th particle. The corresponding distribution functional $P[\rho,t]$ of density field $\rho({\bf r},t)$ is defined by
\begin{equation}
P[\rho,t]=\left<
\prod_{{\bf r}}\delta\left[
\widehat{\rho}_N({\bf r},t)-\rho({\bf r},t)
\right]
\right>,
\label{p def}
\end{equation}
where $\left<\mathcal{O}\right>$ signifies the noise-averaging operation for $\mathcal{O}$ in the overdamped dynamics.

As detailed in Appendix A, $P[\rho,t]$ satisfies the Fokker-Planck equation given by eq. (\ref{appendix fokker}). It follows from the stationary condition $\partial P_{\mathrm{st}}[\rho]/\partial t=0$ on the Fokker-Planck equation that the distribution functional in a steady state, $P_{\mathrm{st}}[\rho]$, is determined by the free-energy functional $\mathcal{A}[\rho]$ of a given density field $\rho({\bf r},t)$:
\begin{flalign}
&P_{\mathrm{st}}[\rho]
=\frac{e^{-\mathcal{A}[\rho]}}{\int
\mathcal{\mathcal{D}}\rho\,e^{-\mathcal{A}[\rho]}}.
\label{pst}
\end{flalign}
We can evaluate the constrained free-energy functional $\mathcal{A}[\rho]$ by introducing a fluctuating potential field $\phi$ as follows (see Appendix A for the derivation of eqs. (\ref{a rho def}) to (\ref{f first})):
\begin{flalign}
e^{-\mathcal{A}[\rho]}&=\int
D\phi\,e^{-F[\rho,\phi]}\Delta[\rho],
\label{a rho def}
\end{flalign}
where $\Delta[\rho]$ denotes the constraint due to the canonical ensemble:
\begin{eqnarray}
\Delta[\rho]=
\left\{
\begin{array}{l}
1\quad(\int d{\bf r}\rho({\bf r})=N)\\
\\ 
0\quad(\int d{\bf r}\rho({\bf r})\neq N).\\
\end{array}
\right.
\label{delta number}
\end{eqnarray}
The functional $F[\rho,\phi]$ in the exponent of eq. (\ref{a rho def}) is defined using the grand potential as follows:
\begin{flalign}
&e^{-F[\rho,\phi]+\int d{\bf r}\rho({\bf r})\mu}\nonumber\\
&=\mathrm{Tr}\prod_i e^{\mu+i\phi({\bf r}_i)-\psi_{\mathrm{dft}}({\bf r}_i)}
\prod_{i,j}e^{-v({\bf r}_i-{\bf r}_j)}\nonumber\\
&\qquad\qquad\qquad\qquad\times
e^{\int d{\bf r}\rho({\bf r})\left\{\psi_{\mathrm{dft}}({\bf r})-i\phi({\bf r})\right\}}\nonumber\\
&=e^{-\Omega\{\psi_{\mathrm{dft}}-i\phi\}+\int d{\bf r}\rho({\bf r})\left\{\psi_{\mathrm{dft}}({\bf r})-i\phi({\bf r})\right\}},
\label{f rhophi def}
\end{flalign}
where $\mathrm{Tr}\equiv\sum_{N=0}^{\infty}\frac{1}{N!}\int d{\bf r}\,_1\cdots\int d{\bf r}\,_N$, $\mu$ denotes the equilibrium chemical potential, $v({\bf r})$ the original interaction potential including the hard sphere potential and Lennard-Jones potential, and $\Omega[\psi]$ the grand potential in the presence of an external field $\psi({\bf r})$. Here it is noted that all of the energetic quantities used in this paper (e.g., $\mu$, $v({\bf r})$ and $\Omega[\psi]$) are given in the $k_BT$--unit.

As clearly seen from Appendix A, the fluctuating potential field $\phi({\bf r})$ in eq. (\ref{f rhophi def}) traces back to the auxiliary field for the Fourier transform of the Dirac delta functional $\prod_{{\bf r}}\delta\left[ \widehat{\rho}_N({\bf r},t)-\rho({\bf r},t) \right]$ \cite{frusawa1,frusawa2,frusawa3}. Also, the functional $F[\rho,\phi\equiv 0]$ in the absence of the $\phi$--field corresponds to the {\itshape intrinsic} Helmholtz free energy in the presence of the external field $\psi_{\mathrm{dft}}({\bf r})$. Therefore, the following relation holds:
\begin{flalign}
\frac{\delta F[\rho,0]}{\delta\rho({\bf r})}
&=\mu-\psi_{\mathrm{dft}}({\bf r}),
\label{f first}
\end{flalign}
according to the equilibrium DFT \cite{evans,singh,likos,lut1}.

\subsection{Non-equilibrium excess chemical potential $\lambda_{\mathrm{ex}}[\rho]$}
The Fokker-Planck equation for $P[\rho,t]$ is equivalent to the stochastic DFT equation, or the Dean-Kawasaki equation \cite{witt}, which is given by
\begin{flalign}
\frac{\partial\rho({\bf r},t)}{\partial t}
&=\nabla\cdot\mathcal{D}\rho\nabla\lambda_{\mathrm{ex}}[\rho]+\zeta[\rho,\overrightarrow{\eta}],
\label{dean kawasaki}\\
\lambda_{\mathrm{ex}}[\rho]&=\frac{\delta
\mathcal{A}[\rho]}{\delta\rho},
\label{lambda def}
\end{flalign}
where $\zeta[\rho,\overrightarrow{\eta}]$ can be expressed as $\zeta[\rho,\overrightarrow{\eta}]=-\nabla\cdot\sqrt{2\mathcal{D}\rho({\bf r},t)}\overrightarrow{\eta}({\bf r},t)$ using the bare diffusion constant $\mathcal{D}$ and the vectorial white noise field $\overrightarrow{\eta}({\bf x},t)$ defined by the correlation $\left<\eta_l({\bf r},t)\eta_m({\bf r}',t')\right>=\delta_{lm}\delta({\bf r}-{\bf r}')\delta(t-t')$, and $\lambda_{\mathrm{ex}}[\rho]$ will be referred to as the non-equilibrium excess chemical potential.

Combining eqs. (\ref{a rho def}) and (\ref{lambda def}), we have
\begin{flalign}
\lambda_{\mathrm{ex}}[\rho]=\frac{\int D\phi\,\frac{\delta
F[\rho,\phi]}{\delta\rho}\,e^{-F[\rho,\phi]}\Delta[\rho]}{\int
D\phi\,e^{-F[\rho,\phi]}\Delta[\rho]},
\label{lambda def2}
\end{flalign}
which further reads
\begin{flalign}
\lambda_{\mathrm{ex}}[\rho]
&=\overline{\lambda_{\mathrm{ex}}[\rho,\phi]}\nonumber\\
&\equiv\frac{\int
D\phi\,\lambda_{\mathrm{ex}}[\rho,\phi]\,e^{-\Delta
F_{\mathrm{dft}}[\rho,\phi]}}{\int D\phi\,e^{-\Delta
F_{\mathrm{dft}}[\rho,\phi]}},
\label{lambda av}\\
\lambda_{\mathrm{ex}}[\rho,\phi]&=\frac{\delta
F[\rho,0]}{\delta\rho}
+\frac{\delta \Delta
F_{\mathrm{dft}}[\rho,\phi]}{\delta\rho}-\lambda_N,
\label{lambda rhophi split}
\end{flalign}
where $\Delta F_{\mathrm{dft}}[\rho,\phi]$ signifies the free-energy difference between $F[\rho,\phi]$ and $F[\rho,0]$, and the Lagrange multiplier $\lambda_N$ enforces the number constraint $\int d{\bf r}\rho({\bf r})=N$ and is reduced to the chemical potential $\mu$ when considering the equilibrium DFT (see Sec. IIIC).

In the Gaussian approximation of the $\phi$--field, we have
\begin{flalign}
\Delta
F_{\mathrm{dft}}[\rho,\phi]&=F[\rho,\phi]-F[\rho,0]\nonumber\\
&=\frac{1}{2}\iint d{\bf r}d{\bf r}'\phi({\bf r})w^{-1}({\bf r}-{\bf r}')\phi({\bf r}'),
\label{delta f w}\\
\frac{\delta^2 F[\rho,0]}{\delta\rho({\bf r})\delta\rho({\bf r}')}
&=\overline{\phi({\bf r})\phi({\bf r}')}=w({\bf r}-{\bf r}'),
\label{f second w}
\end{flalign}
using the density-density correlation function $w^{-1}({\bf r}-{\bf r}')$ (see Appendix A6 for the details). For the concrete representation of the above propagator $w({\bf r})$, we define the DCF $c({\bf r})$ and the total correlation function $h({\bf r})$, based on the equilibrium DFT. The propagator $w({\bf r})$ is expressed as
\begin{flalign}
w({\bf r}-{\bf r}')&=\frac{\delta({\bf r}-{\bf r}')}{\rho({\bf r})}-c({\bf r}-{\bf r}'),
\label{w c}\\
w^{-1}({\bf r}-{\bf r}')&=\rho({\bf r})\left\{
\delta({\bf r}-{\bf r}')+h({\bf r}-{\bf r}')\rho({\bf r}')\right\}.
\label{w-1 h}
\end{flalign}
Equations (\ref{w c}) and (\ref{w-1 h}) manifest that the equilibrium DFT is incorporated into the stochastic DFT.

It follows from eqs. (\ref{lambda def2}) to (\ref{lambda rhophi split}) that
\begin{flalign}
\lambda_{\mathrm{ex}}[\rho]=
\frac{\delta F[\rho,0]}{\delta\rho}
+\overline{\frac{\delta \Delta
F_{\mathrm{dft}}[\rho,\phi]}{\delta\rho}}-\lambda_N,
\label{delta lambda}
\end{flalign}
giving
\begin{flalign}
\nabla\lambda_{\mathrm{ex}}[\rho]&=\nabla\overline{\frac{\delta
F[\rho,\phi]}{\delta\rho}}=\nabla\frac{\delta
F[\rho,0]}{\delta\rho}
+\nabla\overline{\frac{\delta \Delta
F_{\mathrm{dft}}[\rho,\phi]}{\delta\rho}}
\label{lambda grad}
\end{flalign}
because of $|\nabla\lambda_N|=0$. In Sec. V, we will evaluate the second term on the right hand side (rhs) of eq. (\ref{delta lambda}), $\overline{\delta \Delta F_{\mathrm{dft}}[\rho,\phi]/\delta\rho}$, following the above Gaussian approximation, whereas the strong-coupling approximation developed for the evaluation of $F[\rho,\phi]$ will be presented in Sec. VII.

\subsection{Comparison with the deterministic DFT equation}
It has been proved in various ways that the stochastic DFT equation (\ref{dean kawasaki}) is converted into the deterministic DFT equation \cite{witt} when neglecting the additional free-energy functional $\Delta F_{\mathrm{dft}}[\rho,\phi]$: eq. (\ref{lambda grad}) reduces to
\begin{flalign}
\nabla\lambda_{\mathrm{ex}}[\rho]&=\nabla\frac{\delta
F[\rho,0]}{\delta\rho}
\label{lambda grad2}
\end{flalign}
when the last noise term on rhs of eq. (\ref{dean kawasaki}) disappears.

Going back to eqs. (\ref{f second w}) and (\ref{w c}), we find that the Ramakrishnan-Yussouff functional \cite{ry} of the intrinsic Helmholtz free energy $F[\rho,0]$ is of the following form:
\begin{flalign}
&F[\rho,0]\nonumber\\
&=F[n,0]
-\frac{1}{2}\iint d{\bf r}d{\bf r}'\Delta\rho({\bf r})c({\bf r}-{\bf r}')\Delta\rho({\bf r}')
+\Delta F_{\mathrm{id}}[\rho],\nonumber\\
&F_{\mathrm{id}}[\rho]
=\int d{\bf r}\,\rho({\bf r})\left\{\ln\rho({\bf r})-1\right\},
\label{ry functional}
\end{flalign}
where $\Delta\rho\equiv\rho-n$, $\Delta F_{\mathrm{id}}[\rho]\equiv F_{\mathrm{id}}[\rho]-F_{\mathrm{id}}[n]$, and $n=N/V$ denotes the uniform mean density with $V$ being the system volume. Combination of eqs. (\ref{f first}) and (\ref{ry functional}) provides
\begin{flalign}
\mu=\ln\rho({\bf r})+\psi_{\mathrm{dft}}({\bf r})-\int d{\bf r}'c({\bf r}-{\bf r}')\Delta\rho({\bf r}'),
\label{dft chem}
\end{flalign}
which reads
\begin{flalign}
\rho({\bf r})=e^{\mu-\psi_{\mathrm{dft}}({\bf r})+\int d{\bf r}'c({\bf r}-{\bf r}')\Delta\rho({\bf r}')},
\label{prescribed density}
\end{flalign}
stating that the conventional relation of the equilibrium DFT for a prescribed density $\rho({\bf r})$ is satisfied by adjusting the external potential $\psi_{\mathrm{dft}}({\bf r})$ (see also eq. (\ref{appendix prescribe})).

We obtain from plugging eq. (\ref{ry functional}) into eq. (\ref{lambda grad2}) the deterministic DFT equation with the use of the Ramakrishnan-Yussouff functional:
\begin{flalign}
&\frac{\partial\rho({\bf r},t)}{\partial t}
=\nabla\cdot\mathcal{D}\rho\nabla\frac{\delta
F[\rho,0]}{\delta\rho}\nonumber\\
&=\mathcal{D}\nabla^2\rho({\bf r},t)-\nabla\cdot\mathcal{D}\rho({\bf r},t)\int d{\bf r}'\nabla
c({\bf r}-{\bf r}')\Delta\rho({\bf r}',t).
\label{dean kawasaki2}
\end{flalign}
Comparison between eqs. (\ref{dean kawasaki}) and (\ref{dean kawasaki2}), or between eqs. (\ref{lambda grad}) and (\ref{lambda grad2}), indicates the difference between the stochastic and deterministic DFT equations.

\subsection{Defining metastable states based on the stochastic DFT}
Before considering the metastability condition for the stochastic DFT equation (\ref{dean kawasaki}), we connect a metastable distribution $\rho^*_{\mathrm{dft}}({\bf r})$ determined by the equilibrium DFT with the deterministic DFT equation (\ref{dean kawasaki2}). In the absence of external field (i.e., $\psi_{\mathrm{dft}}\equiv 0$), eq. (\ref{f first}) reduces to the metastability condition for the equilibrium DFT:
\begin{flalign}
\left.
\frac{\delta F[\rho,0]}{\delta\rho({\bf r})}\right|_{\rho=\rho^*_{\mathrm{dft}}}
&=\mu,
\label{dft metastable}
\end{flalign}
where $F[\rho^*_{\mathrm{dft}},0]$ becomes equal to the intrinsic Helmholtz free energy defined in equilibrium. Correspondingly, eq. (\ref{dft chem}) leads to
\begin{flalign}
\mu=\ln\rho^*_{\mathrm{dft}}({\bf r})-\int d{\bf r}'c({\bf r}-{\bf r}')\Delta\rho^*_{\mathrm{dft}}({\bf r}').
\label{dft chem2}
\end{flalign}
The non-equilibrium excess chemical potential $\lambda_{\mathrm{ex}}[\rho]$ should disappear at $\rho^*_{\mathrm{dft}}$:
\begin{flalign}
\lambda_{\mathrm{ex}}[\rho^*_{\mathrm{dft}}]
&=\left.
\frac{\delta
F[\rho,0]}{\delta\rho}\right|_{\rho=\rho^*_{\mathrm{dft}}}-\lambda_N\nonumber\\
&=\mu-\lambda_N=0,
\label{eq excess}
\end{flalign}
implying that the Lagrange multiplier $\lambda_N$ is correctly identified with the equilibrium chemical potential $\mu$ at $\rho^*_{\mathrm{dft}}$, as mentioned above.

The difference between the equilibrium and stochastic DFTs can be clearly seen from plugging eq. (\ref{dft metastable}) into eqs. (\ref{dean kawasaki}) and (\ref{dean kawasaki2}). On the one hand, the deterministic DFT equation (\ref{dean kawasaki2}) ensures that eq. (\ref{dft metastable}) is a steady-state condition: we have $\partial\rho^*_{\mathrm{dft}}/\partial t=0$ because the rhs of eq. (\ref{dean kawasaki2}) vanishes due to $|\nabla\delta F[\rho^*_{\mathrm{dft}},0]/\delta\rho^*_{\mathrm{dft}}|=|\nabla\mu|=0$. On the other hand, the stochastic DFT equation (\ref{dean kawasaki}) for $\rho^*_{\mathrm{dft}}({\bf r},t)$ becomes
\begin{flalign}
\frac{\partial\rho^*_{\mathrm{dft}}({\bf r},t)}{\partial t}
&=\nabla\cdot\mathcal{D}\rho^*_{\mathrm{dft}}\nabla\overline{\frac{\delta\Delta
F_{\mathrm{dft}}[\rho^*_{\mathrm{dft}},\phi]}{\delta\rho^*_{\mathrm{dft}}}}+\zeta[\rho^*_{\mathrm{dft}},\overrightarrow{\eta}],
\label{dean kawasaki remain}
\end{flalign}
revealing that, in genera, $\rho^*_{\mathrm{dft}}({\bf r},t)$ is not a steady-state distribution in terms of the stochastic DFT.

Meanwhile, the metastability condition for the stochastic DFT equation (\ref{dean kawasaki}) is that the metastable excess chemical potential $\lambda_{\mathrm{ex}}[\rho^*]$ given by eq. (\ref{delta lambda}) does not necessarily vanish but has a spatially constant value $\lambda_{\mathrm{ex}}^*$:
\begin{flalign}
\left.\frac{\delta
\mathcal{A}[\rho]}{\delta\rho}\right|_{\rho=\rho^*}
=\lambda_{\mathrm{ex}}^*.
\label{metastable}
\end{flalign}
The first term on the rhs of eq. (\ref{dean kawasaki}) disappears when eq. (\ref{metastable}) is satisfied, yielding
\begin{equation}
\left<
\frac{\partial \rho^*({\bf r},t)}{\partial t}\right>
=\left<\zeta[\rho^*,\overrightarrow{\eta}]\right>=0,
\label{metastable dk}
\end{equation}
on noise-averaging. A previous study based on the stochastic thermodynamics has shown that the heat dissipated into the reservoir is negligible on average when satisfying eq. (\ref{metastable}) or eq. (\ref{metastable dk}) \cite{frusawa1} .

In this study, we thus adopt the metastability condition (\ref{metastable}) based on the stochastic DFT, instead of eq. (\ref{dft metastable}).

\begin{table*}
\caption{Four types of hyperuniform and non-hyperuniform systems. We investigate the type N1 of non-hyperuniform systems.}
\centering
\begin{tabular}{c|cc|c}
&&&\\
&\multicolumn{2}{|c|}{Characterization of the DCF}&\\[6pt]
System &\quad Power-law decay$\quad$&\quad Magnitude at zero
separation$\quad$&\quad Type\\[6pt]
\hline\hline
&&\\
\multirow{2}{*}{Hyperuniform} &Complete&---&H1\\[6pt]
 &Incomplete&Divergent&H2\\[6pt]
\hline
&&\\
\multirow{2}{*}{Non-hyperuniform} &\underline{\bf Incomplete}&\underline{\bf Finite}&\fbox{\bf N1}\\[6pt]
&Absent &Finite&N2
\end{tabular}
\end{table*}

\subsection{The zero-wavevector structure factor $S^*(0)$ in a metastable state defined by eq. (\ref{metastable})}
It has been demonstrated near and at jamming that the structure factor $S^*(k)$ in a metastable state can be written as
\begin{flalign}
S^*(k)&=\frac{1}{N}\left<\rho^*(k)\right>\left<\rho^*(-k)\right>,
\label{s def}
\end{flalign}
because the structure factor in a frozen state mainly arises from the configurational part which is associated with the averaged positions of arrested particles \cite{ikeda non1}. Equations (\ref{eq excess}), (\ref{metastable}) and (\ref{s def}) imply that $S^*(0)$ is obtained from the metastable chemical potential,
\begin{equation}
\lambda^*=\lambda_{\mathrm{ex}}^*+\lambda_N,
\label{metastable chem}
\end{equation}
in a similar manner to the equilibrium DFT as follows:
\begin{flalign}
\frac{1}{S^*(0)}
&=n\left.
\frac{\delta\lambda^*}{\delta\rho^*}\right|_{\rho^*=n}
\label{s lambda}\\
&=1-n\int d{\bf r}\,c^*({\bf r})\nonumber\\
&=-c^*(0)-n\int_{r\geq\sigma}4\pi r^2dr\,c^*({\bf r})\nonumber\\&\qquad-n\int_{r\geq\sigma}4\pi
r^2dr\,c^*({\bf r})h^*({\bf r})\nonumber\\
&\approx-c^*(0)-n\int_{r\geq\sigma}4\pi r^2dr\,c^*({\bf r}),
\label{s approx}
\end{flalign}
where the metastable DCF $c^*({\bf r})$ is defined by eq. (\ref{s lambda}) using the metastable chemical potential $\lambda^*$ and the approximate expression given in the last line of eq. (\ref{s approx}) is obtained from the Ornstein-Zernike equation regarding $c^*({\bf r})$ at zero separation $r=0$ (see Appendix B1 for the detailed derivation).

The equilibrium DFT, on the other hand, provides the metastable density distribution $\rho^*_{\mathrm{dft}}({\bf r})$ determined by eq. (\ref{dft metastable}). It follows that the metastable structure factor $S^*(k)$ reads
\begin{flalign}
S^*(k)&=\frac{1}{N}\rho^*_{\mathrm{dft}}(k)\rho^*_{\mathrm{dft}}(-k),
\label{s eqDFT}
\end{flalign}
with $\left<\rho^*(k)\right>$ in eq. (\ref{s def}) being replaced by $\rho^*_{\mathrm{dft}}(k)$. Then, we obtain from eqs. (\ref{ry functional}) and (\ref{dft metastable})
\begin{flalign}
\frac{1}{S^*(0)}
&=n\left.
\frac{\delta\mu}{\delta\rho_{\mathrm{dft}}^*}\right|_{\rho_{\mathrm{dft}}^*=n}\nonumber\\
&=1-n\int d{\bf r}\,c({\bf r}),
\label{s eq approx}
\end{flalign}
confirming that the DCF $c({\bf r})$ determines the zero-wavevector structure factor.

\section{Our aim: non-hyperuniform states on target}
Table 1 classifies hyperuniform and non-hyperuniform systems into four types for clarifying the non-hyperuniform state to be addressed hereafter. The type H1 in Table 1 signifies a hyperuniform state without requirement for $c^*(0)$ because eq. (\ref{hyper dcf}) is completely satisfied (i.e., $L_c\rightarrow\infty$).

Despite the finiteness of long-range nature, eq. (\ref{s approx}) still predicts that the hyperuniformity of type H2 is necessarily observed near and at jamming unless the zero-separation divergence of $c^*(0)$ is avoided. This is because $1/S^*(0)$ diverges due to either the long-range nature or the divergent behavior at zero separation, as found from combining eqs. (\ref{hyper dcf}) and (\ref{s approx}).

To summarize, there are two requirements on the non-hyperuniform DCF $c^*({\bf r})$ of type N1 as follows:
\begin{itemize}
\item[(i)] {\itshape Finiteness of the long-range nature}.--- The non-hyperuniformity requires a drop in the long-ranged DCF for $r>L_c$. Namely, the first requirement is that $c^*({\bf r})$ must decay rapidly to zero for $r>L_c$ \cite{chieco,silbert,olsson,ikeda non1,ikeda non2}; otherwise, the second term on the rhs of eq. (\ref{s approx}) is divergent.
\item[(ii)] {\itshape Short-range cutoff}.--- As seen from the first term on the rhs of eq. (\ref{s approx}), the metastable DCF at zero separation (i.e., $c^*(0)$) must have a finite value even as the densely packed systems approach jamming, which is the second requirement.
\end{itemize}
Equation (\ref{s approx}) reveals that the zero-wavevector structure factor never vanishes without meeting both of the above requirements. Nevertheless, exclusive attention in previous studies \cite{chieco,silbert,olsson,ikeda non1,ikeda non2} has been paid to the former requirement, and the short-range cutoff of the metastable DCF (the second requirement (ii)) remains to be investigated.

In reality, the zero-separation DCF tends to have an extremely large value near freezing in repulsive sphere systems; for instance, the Percus-Yevick approximation of hard sphere fluids provides \cite{py}
\begin{equation}
-k_BT c^*(0)=\frac{\partial P}{\partial n},
\label{py}
\end{equation}  
suggesting the divergent behavior of $-c^*(0)$ in a frozen state.

Thus, we focus on the emergence of type N1 when investigating the degradation of hyperuniformity. To be more specific, we show theoretically that the non-hyperuniformity of type N1 satisfies
\begin{eqnarray}
c^*({\bf r})\sim
\left\{
\begin{array}{l}
\mathcal{C}\quad(r=0)\\
\\ 
\frac{1}{r^2}\quad(\sigma<r\leq L_c),\\
\end{array}
\right.
\label{n1 dcf}
\end{eqnarray}
though the hyperuniformity of type H2 is incorporated into the equilibrium DFT as input:
\begin{eqnarray}
c({\bf r})\sim
\left\{
\begin{array}{l}
\frac{1}{r}\quad(0\leq r\leq\sigma)\\
\\ 
\frac{1}{r^2}\quad(\sigma<r\leq L_c),\\
\end{array}
\right.
\label{h2 dcf}
\end{eqnarray}
where $c({\bf r})$ is different from the completely hyperuniform DCF in that $L_c$ is supposed to have a finite value. Following the equilibrium DFT, eqs. (\ref{s eq approx}) and (\ref{h2 dcf}) lead to $S^*(0)\sim -c(0)>10^4$ despite the finiteness of the long-range power-law decay, which is the hyperuniformity of type H2.

We are now ready to address the issues on the non-hyperuniformity of type N1. In what follows, we present a preliminary result obtained in the Gaussian approximation for comparing the stochastic and equilibrium DFTs, and subsequently {\itshape prove in the strong-coupling approximation of the stochastic DFT that eq. (\ref{h2 dcf}) transforms to eq. (\ref{n1 dcf}) as a result of the ensemble average over the fluctuating $\phi$--field (see also eq. (\ref{lambda av}))}.

\section{Gaussian approximation of the stochastic DFT}
In the first place, we investigate the free-energy functional difference between the stochastic and equilibrium DFTs when performing the Gaussian approximation given by eq. (\ref{delta f w}). In the Gaussian approximation, eq. (\ref{delta lambda}) reduces to
\begin{flalign}
&\lambda[\rho]\equiv
\lambda_{\mathrm{ex}}[\rho]+\lambda_N\nonumber\\
&=\frac{\delta F[\rho,0]}{\delta\rho({\bf r})}
+\frac{1}{2}\overline{\frac{\delta}{\delta\rho}\iint d{\bf r}d{\bf r}'w^{-1}({\bf r}-{\bf r}')
\phi({\bf r})\phi{\bf r}')}.
\label{lambda gaussian}
\end{flalign}
As seen from Appendix A7 for the details, we have
\begin{flalign}
&\overline{\frac{\delta}{\delta\rho}\iint d{\bf r}d{\bf r}'w^{-1}({\bf r}-{\bf r}')
\phi({\bf r})\phi{\bf r}')}\nonumber\\
&=\frac{\frac{\delta}{\delta\rho}\iint d{\bf r}d{\bf r}'w^{-1}({\bf r}-{\bf r}')
\int D\phi\,\phi({\bf r})\phi({\bf r}')e^{-\Delta F_{\mathrm{dft}}[\rho,\phi]}}
{\int D\phi\,e^{-\Delta F_{\mathrm{dft}}[\rho,\phi]}}
\nonumber\\
&\approx c(0)-h(0).
\label{phiphi transform}
\end{flalign}
Combining eqs. (\ref{ry functional}), (\ref{metastable chem}) and (\ref{phiphi transform}), eq. (\ref{lambda gaussian}) reads
\begin{flalign}
&\lambda^*=\lambda[\rho^*]=\lambda^*_{\mathrm{ex}}+\lambda_N\nonumber\\
&=\ln\rho^*({\bf r})-\int d{\bf r}'c({\bf r}-{\bf r}')\Delta\rho^*({\bf r}')+\frac{1}{2}\left\{c(0)-h(0)\right\},
\label{lambda gaussian result}
\end{flalign}
in a metastable state.
We find from eqs. (\ref{s lambda}) and (\ref{lambda gaussian result})
\begin{flalign}
\frac{1}{S^*(0)}
&=1-n\int d{\bf r}\,c({\bf r})
\label{s gaussian}
\end{flalign}
while neglecting $\delta c(0)/\delta\rho$ , or the triplet DCF. Comparison between eqs. (\ref{s eq approx}) and (\ref{s gaussian}) confirms that no degradation of hyperuniformity is induced by Gaussian potential fluctuations.

To see the correspondence with previous results, it is convenient to transform eq. (\ref{lambda gaussian result}) to
\begin{flalign}
\rho^*({\bf r})=e^{\lambda^*
+\int d{\bf r}'c({\bf r}-{\bf r}')\Delta\rho^*({\bf r}')
-\frac{1}{2}\left\{c(0)-h(0)\right\}}.
\label{gaussian density}
\end{flalign}
Equation (\ref{gaussian density}) is, on the one hand, of the same form as the previous results obtained from the Gaussian approximation in various ways when $\lambda^*=\mu$ \cite{frydel,frusawa4}. On the other hand, comparison between eq. (\ref{prescribed density}) with $\psi_{\mathrm{dft}}\equiv 0$ and eq. (\ref{gaussian density}) indicates that eq. (\ref{gaussian density}) is identical to the self-consistent equation of $\rho^*({\bf r})$ conventionally used in the equilibrium DFT when
\begin{flalign}
\lambda^*&=\mu+\frac{1}{2}\left\{c(0)-h(0)\right\}
\label{correspondence1}
\end{flalign}
when $\lambda_N=\mu$ and $\lambda_{\mathrm{ex}}=\left\{c(0)-h(0)\right\}/2$.
Equation (\ref{correspondence1}) reveals that stochastic fluctuations create an additional contribution, the second term on the rhs of eq. (\ref{correspondence1}), to the equilibrium chemical potential $\mu$.

\begin{figure*}[hbtp]
\begin{center}
	\includegraphics[
	width=15cm
]{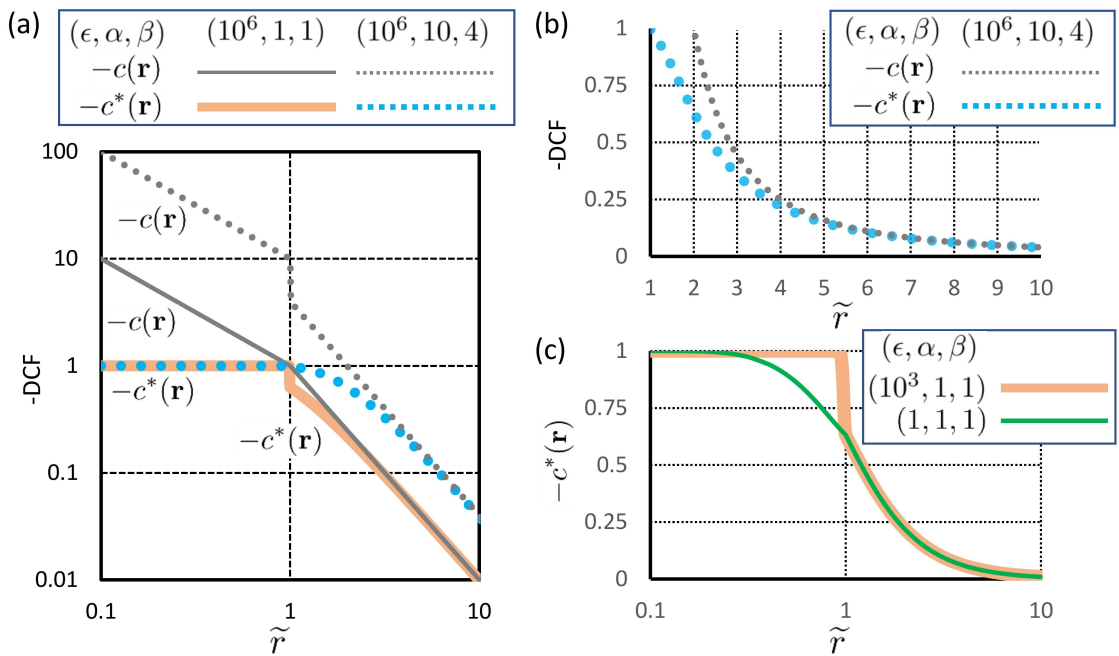}
\end{center}
\caption{
Comparison between the metastable DCF $c^*({\bf r})$ given by eq. (\ref{c result}) and the hyperuniform DCF $c({\bf r})$ expressed by eq. (\ref{fig1 dcf}) for the parameter sets of $(\epsilon, \alpha, \beta)$ as follows: while we need to fix two parameters, $\alpha$ and $\beta$, for representing the expression (\ref{fig1 dcf}) of $c({\bf r})$, it is necessary to set not only $\alpha$ and $\beta$, but also the parameter $\epsilon$ of the original interaction potential $v({\bf r})$ given by eq. (\ref{harmonic}) for showing the obtained form (\ref{c result}) of $c^*({\bf r})$. (a) A log-log plot of $c({\bf r})$ and $c^*({\bf r})$ which are depicted using the parameter sets as follows: $(\epsilon,\alpha,\beta)=(10^6,10,4)$ and $(10^6,1,1)$. (b) A linear plot for comparing $c({\bf r})$ and $c^*({\bf r})$ with the parameter set of $(\epsilon,\alpha)=(10^6,10,4)$ in more detail. (c) A semi-log plot of $c^*({\bf r})$ when $\epsilon$ is decreased from $10^6$ to either $10^3$ or $1$. The metastable DCF $c^*({\bf r})$ is softened with the decrease of $\epsilon$ in eq. (\ref{harmonic}).
}
\end{figure*}

\begin{figure}[hbtp]
\begin{center}
	\includegraphics[
	width=8cm
]{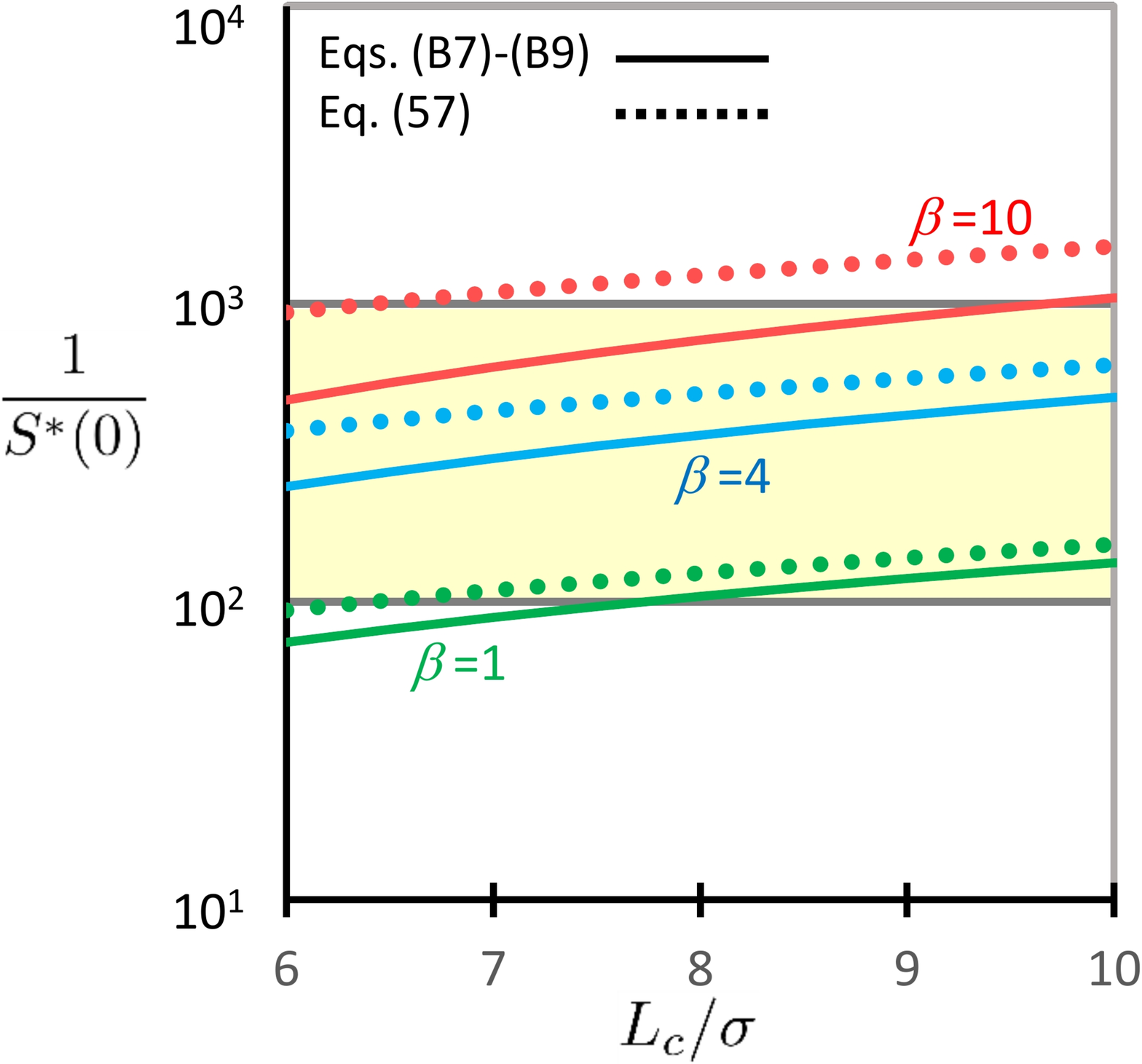}
\end{center}
\caption{
The last expression in eq. (\ref{s approx}) can be calculated analytically when using eqs. (\ref{c result}) and (\ref{fig1 dcf}). The three solid lines depict the analytical result (\ref{appendix s0 exact}) with eqs. (\ref{appendix s0 i1}) and (\ref{appendix erf}), or the precise results of the zero-wavevector structure factor $S^*(0)$, for $\beta=1$, 4 and 10 at $f_{\mathrm{v}}=0.65$. For comparison, the dotted lines representing the approximate form (\ref{approximate structure}) are also drawn for the same parameter sets: $\beta=1$, 4 and 10 at $f_{\mathrm{v}}=0.65$. The yellow area corresponds to the non-hyperuniform range of $1/S^*(0)$ which is given by either eq. (\ref{nonhyper result}) or eq. (\ref{meta s comp}).
}
\end{figure}

\section{Main results and comparison with simulation results}
To go beyond the Gaussian approximation, we need to explore an expansion method adequate for strongly-correlated sphere systems near and at jamming. One candidate is the virial-type expansion that has proven to be applicable to inhomogeneous ionic fluids at strong coupling \cite{netz}. In the next section, we will verify that the virial-type expansion can apply also to the evaluation of $\lambda_{\mathrm{ex}}[\rho]$ given by eq. (\ref{lambda av}), hence yielding the metastable DCF $c^*({\bf r})$ other than $c({\bf r})$.

In this section, the obtained form of the metastable DCF $c^*({\bf r})$, which satisfies the relation (\ref{n1 dcf}), is presented in advance (Sec. VIA). Subsequently, the calculated value of $S^*(0)$ is compared with simulation results (i.e., eq. (\ref{nonhyper result})) on the non-hyperuniform structure factor at jamming (Sec. VIB).

\subsection{Typical behaviors of the metastable DCF $c^*({\bf r})$}
As proved in the next section, the development of the strong-coupling expansion method, or the $1/\gamma$ expansion method, allows us to find the following form of the metastable chemical potential $\lambda^*$:
\begin{flalign}
&\lambda^*
=\ln\rho^*({\bf r})+\frac{w(0)}{2}
-\int d{\bf r}'c^*({\bf r}-{\bf r}')\rho^*({\bf r}'),
\label{lambda result}\\
&-c^*({\bf r}-{\bf r}')=
1-e^{-v({\bf r}-{\bf r}')-w({\bf r}-{\bf r}')}.
\label{c result}
\end{flalign}
Since the relation $w(0)\gg 1$ holds at jamming, eq. (\ref{c result}) leads to
\begin{flalign}
-c^*(0)=1,
\label{zero separation result}
\end{flalign}
regardless of the repulsive potential form of $v({\bf r})$. Equation (\ref{zero separation result}) reveals that, in general, the metastable DCF $c^*({\bf r})$ given by (\ref{c result}) satisfies the second requirement (or eq. (\ref{n1 dcf})) for the non-hyperuniformity (see the requirement (ii) in Sec. IV). In particular for hard spheres, the resulting form (\ref{c result}) reads
\begin{eqnarray}
-c^*({\bf r})=
\left\{
\begin{array}{l}
1\quad(\widetilde{r}\leq 1)\\
\\
-c({\bf r})\quad(\widetilde{r}\gg 1),\\
\end{array}
\right.
\label{main approx}
\end{eqnarray}
where $\widetilde{r}\equiv r/\sigma$. Equation (\ref{main approx}) meets the above non-hyperuniformity requirements given by eq. (\ref{n1 dcf}) with $\mathcal{C}=-1$.

While the short-range cutoff is seen in the third term on the rhs of eq. (\ref{lambda result}), the second term on the rhs of eq. (\ref{lambda result}) corresponds to the effective self-energy which is divergent due to the power-law behavior expressed by eq. (\ref{h2 dcf}). This implies that the effective self-energy term ($=w(0)/2$) offsets the decrease in the interaction contribution due to the short-range cutoff.

Thus, we have obtained various forms of the chemical potential given by eqs. (\ref{dft chem2}), (\ref{lambda gaussian result}) and (\ref{lambda result}) from the equilibrium DFT, the stochastic DFT in the Gaussian approximation, and the stochastic DFT in the strong-coupling approximation, respectively.
The above discussions suggest that different results of the hyperuniform and non-hyperuniform chemical potentials (i.e., eqs. (\ref{dft chem2}) and (\ref{lambda result})) are compatible with each other in terms of the absolute values.

In Fig. 1, comparison is made between the $\widetilde{r}$-dependencies of $-c({\bf r})$ and $-c^*({\bf r})$ for the repulsive harmonic potential given by
\begin{equation}
v({\bf r})=\epsilon\left(1-\widetilde{r}\right)^2\Theta\left(1-\widetilde{r}\right),
\label{harmonic}
\end{equation}
where $\epsilon$ controls the interaction strength in the $k_BT$-unit and $\Theta(x)$ is the Heaviside step function. It is supposed in Fig. 1 that $-c({\bf r})$ is of the following form:
\begin{eqnarray}
-c({\bf r})=
\left\{
\begin{array}{l}
\frac{\alpha}{\;\widetilde{r}\;}\quad(\widetilde{r}\leq 1)\\
\\
\frac{\beta}{\;\widetilde{r}^2\;}\quad(1<\widetilde{r}\leq\frac{L_c}{\sigma})\\
\\
0\qquad(\widetilde{r}>\frac{L_c}{\sigma}).
\end{array}
\right.
\label{fig1 dcf}
\end{eqnarray}
To be noted, eq. (\ref{fig1 dcf}) does not include the delta function $-\frac{1}{4\phi}\delta(\widetilde{r}-1)$ due to the isostaticity, a significant negative contribution to $-c({\bf r})$ at $\widetilde{r}=1$ \cite{ t dcf}.

Previous simulation studies \cite{ozawa,chieco,silbert,olsson,ikeda non1,ikeda non2} have indicated the parameter ranges of $\epsilon\geq 10^4$, $\alpha\sim10^1$ and $0.1\leq\beta\leq 10^1$ close to jamming. Correspondingly, we consider four sets of parameters in Fig. 1: $(\epsilon,\alpha,\beta)=(10^6,10,4),\>(10^6,1,1),\>(10^3,1,1)$ and $(1,1,1)$. In Fig. 1(a), the hyperuniform and metastable DCFs, $-c({\bf r})$ and $-c^*({\bf r})$, are depicted for two sets of parameters, $(\epsilon,\alpha,\beta)=(10^6,10,4)$ and $(10^6,1,1)$, on a log-log plot. We can see from Fig. 1(a) that the potential value of the metastable DCF saturates to unity irrespective of the short-range behavior of $-c({\bf r})$, and that the short-range deviation of $-c^*({\bf r})$ from $-c({\bf r})$ is larger with the increase of $\alpha$ and $\beta$. A magnified view for $r\geq\sigma$ is shown in Fig. 1(b), allowing us to make a comparison between $-c({\bf r})$ and $-c^*({\bf r})$ for $(\epsilon,\alpha,\beta)=(10^6,10,4)$ in more detail. Figure 1(b) shows that $-c^*({\bf r})$ converges to $-c({\bf r})$ for $r\gg\sigma$ even when there is an obvious difference in the DCFs at $r=2\sigma$ between $-c(r=2\sigma)=\beta/4$ and $-c^*(r=\sigma)=1-e^{-\beta/4}$ for $\beta=4$. Figure 1(c) compares the profiles of $-c^*({\bf r})$ for $\epsilon=10^3$ and 1 with $\alpha$ and $\beta$ being the same value ($\alpha=\beta=1$) on a semi-log plot. This indicates that the metastable DCF inside the sphere (i.e., $-c^*({\bf r})$ for $r\leq\sigma$) is not changed until the interaction strength represented by the parameter $\epsilon$ is reduced considerably (for instance, $\epsilon=1$ in Fig. 1(c)) far from the jamming values of $\epsilon\geq 10^4$.

\subsection{Comparison with simulation results given by eqs. (\ref{nonhyper result}) and (\ref{lc result})}
It follows from eqs. (\ref{s approx}), (\ref{c result}), (\ref{harmonic}) and (\ref{fig1 dcf}) that the approximate form of the zero-wavevector structure factor $S^*(0)$ is determined by both the volume fraction $f_{\mathrm{v}}$ of packed spheres and the cutoff length $L_c$:
\begin{flalign}
\frac{1}{S^*(0)}\approx 24f_{\mathrm{v}}
\beta\left(\frac{L_c}{\sigma}\right)
\label{approximate structure}
\end{flalign}
for $L_c/\sigma\gg 1$; see Appendix B2 for the detailed derivation. A first choice to investigate the type-N1 non-hyperuniformity at jamming is to set that $L_c/\sigma=10$ and $f_{\mathrm{v}}=0.65$, according to the previous simulation results \cite{ozawa,chieco,silbert,olsson,ikeda non1,ikeda non2} of non-hyperuniform harmonic-core sphere systems. Equation (\ref{approximate structure}) then becomes $1/S^*(0)=156\beta$, implying that the relation (\ref{nonhyper result}) applies to the metastable structure factor:
\begin{equation}
10^2<\frac{1}{S^*(0)}<10^3,
\label{meta s comp}
\end{equation}
with $\beta\sim\mathcal{O}[10^{b}]$ being in a reasonable range of $0\leq b< 1$.

For validation of the above evaluation, Fig. 2 provides the dependences of $1/S^*(0)$ on $L_c/\sigma$ in the range of eq. (\ref{lc result}) for $\beta=1$, 4 and 10 with $f_{\mathrm{v}}=0.65$ being used as before. As seen from Fig. 2, comparison between the precise result (see eq. (\ref{appendix s0 exact}) in Appendix B2) and the approximate expression (\ref{approximate structure}) shows that eq. (\ref{approximate structure}) is an acceptable approximation. The precise results depicted by solid lines in Fig. 2 further verify the relation (\ref{meta s comp}) for $1\leq\beta\leq 10$ in the range of eq. (\ref{lc result}) for $L_c/\sigma$. Thus, we find that the metastable DCF $c^*({\bf r})$ given by eq. (\ref{c result}), one of the main results in this study, quantitatively explains previous simulation results on the non-hyperuniformity of type N1.

It is also suggested by Fig. 2 that $\beta\sim 10^{-1}$ leads to $1/S^*(0)<10^2$ as long as the cutoff of $-c^*(0)=1$ holds. This result appears to contradict previous simulation results \cite{t dcf} in hyperuniform hard sphere systems where not only the small value of $\beta\sim 10^{-1}$ but also the existence of $L_c$ in the range of eq. (\ref{lc result}) have been found. At the same time, however, the divergent relation $-c({\bf r})\approx 10/\widetilde{r}\;(\widetilde{r}<1)$ has been verified for the present hyperuniform hard sphere systems \cite{t dcf}. Accordingly, the divergent behavior of the hyperuniform DCF $-c(0)$ at zero-separation ensures the hyperuniform relation (\ref{hyper result}): the relation, $1/S(0)\approx -c(0)>10^4$, holds even when $\beta\sim 10^{-1}$ and $L_c/\sigma\sim 10^1$, which is exactly the hyperuniform state of the type-H2 in Table. 1.

\section{Verification of the main result given by eqs. (\ref{lambda result}) and (\ref{c result}) in the strong-coupling approximation}
There are three steps to verify both the metastable chemical potential $\lambda^*$ and DCF $c^*({\bf r})$ given by eqs. (\ref{lambda result}) and (\ref{c result}), respectively. First, we define the coupling constant $\gamma$ and present the free-energy functionals rescaled by $\gamma$, suggesting the validity of the strong-coupling expansion method, or the density-expansion method at high density (Sec. VIIA). Second, the non-equilibrium chemical potential $\lambda[\rho]$ defined by eqs. (\ref{lambda def2}) to (\ref{lambda rhophi split}) is calculated for non-interacting spheres at strong coupling (Sec. VIIB). Third, we connect the $1/\gamma$ expansion, which is equivalent to the density expansion (or the fugacity expansion \cite{netz}), with the virial-type term expressed by the Mayer function, thereby proving eqs. (\ref{lambda result}) and (\ref{c result}) (Sec. VIIC).

\subsection{Rescaled free-energy functionals}
We introduce the rescaled propagator $\widetilde{w}({\bf r})$ using a coupling constant $\gamma$:
\begin{flalign}
\widetilde{w}({\bf r})&=\frac{w({\bf r})}{\gamma^2},\nonumber\\
\gamma&= e^{\frac{w(0)}{2}}.
\label{rescaled w}
\end{flalign}
Since we consider the type-H2 hyperuniform systems as mentioned before, it is found from eqs. (\ref{w c}) and (\ref{h2 dcf}) that the coupling constant $\gamma$ is approximated by $\gamma\approx e^{-c(0)/2}$ and becomes extremely large near and at jamming.

We aim to develop the $1/\gamma$ expansion method at strong coupling ($\gamma\gg 1$), provided that $\gamma$ is extremely large but is finite. In the next subsection, we will show that the virial-type expansion method, the density-expansion method, can be regarded as the $1/\gamma$ expansion method. Before proceeding, we see the $\gamma$--dependencies of functionals based on the following criteria:
\begin{itemize}
\item[]{\bf Criterion 1}: $\Delta
F_{\mathrm{dft}}\left[\widetilde{\rho},\widetilde{\phi}\right]\sim
\gamma^0$,
\item[]{\bf Criterion 2}: $\int d{\bf r}'\widetilde{w}({\bf r}-{\bf r}')\widetilde{w}^{-1}({\bf r}'-{\bf r}")=\delta({\bf r}-{\bf r}")$.
\label{criteria}
\end{itemize}
The criterion 1 allows us to discern the perturbative terms at strong coupling, in comparison with the rescaled functional $\Delta F_{\mathrm{dft}}\left[\widetilde{\rho},\widetilde{\phi}\right]\sim\gamma^0$, whereas the criterion 2 is equivalent to the Ornstein-Zernike equation \cite{evans,singh,likos,lut1} for rescaled correlation functions, $\widetilde{c}({\bf r})$ and $\widetilde{h}({\bf r})$, that should be defined to satisfy
\begin{flalign}
\widetilde{w}({\bf r}-{\bf r}')&=\frac{\delta({\bf r}-{\bf r}')}{\widetilde{\rho}({\bf r})}-\widetilde{c}({\bf r}-{\bf r}'),
\label{rescaled w c}\\
\widetilde{w}^{-1}({\bf r}-{\bf r}')&=\widetilde{\rho}({\bf r})\left\{
\delta({\bf r}-{\bf r}')+\widetilde{h}({\bf r}-{\bf r}')\widetilde{\rho}({\bf r}')\right\},
\label{rescaled w-1 h}
\end{flalign}
consistently with the original definitions given by eqs. (\ref{w c}) and (\ref{w-1 h}).

It is found from eqs. (\ref{delta f w}) and (\ref{rescaled w}) that the criterion 1 imposes the potential rescaling as follows:
\begin{equation}
\phi({\bf r})=\gamma\widetilde{\phi}({\bf r}).
\label{phi scaling}
\end{equation}
On the other hand, it follows from the criterion 2, or eqs. (\ref{rescaled w}) to (\ref{rescaled w-1 h}), that the correlation functions and the density field are necessarily rescaled as
\begin{flalign}
c({\bf r})&=\gamma^2 \widetilde{c}({\bf r}),\nonumber\\
h({\bf r})&=\gamma^2 \widetilde{h}({\bf r}),\nonumber\\
\rho({\bf r})&=\frac{\widetilde{\rho}({\bf r})}{\gamma^2},
\label{chrho scaling}
\end{flalign}
to satisfy the criterion 2.

Combination of eqs. (\ref{rescaled w}) and (\ref{phi scaling}) transforms eq. (\ref{delta f w}) to
\begin{flalign}
\Delta
F_{\mathrm{dft}}\left[\widetilde{\rho},\widetilde{\phi}\right]=
\frac{1}{2}\iint d{\bf r}d{\bf r}'
\widetilde{\phi}({\bf r})\widetilde{w}^{-1}({\bf r}-{\bf r}')\widetilde{\phi}({\bf r}'),
\label{rescaled delta f}
\end{flalign}
meeting the above criterion 1. Meanwhile, the rescaled form $F\left[\widetilde{\rho},0\right]$ of the Ramakrishnan-Yussouff free energy functional (\ref{ry functional}) is
\begin{flalign}
&F\left[\widetilde{\rho},0\right]\nonumber\\
&=F\left[\widetilde{n},0\right]-\frac{1}{2\gamma^2}\iint d{\bf r}d{\bf r}'\widetilde{\Delta\rho}({\bf r})\widetilde{c}({\bf r}-{\bf r}')\widetilde{\Delta\rho}({\bf r}')
+\Delta F_{\mathrm{id}}\left[\widetilde{\rho}\right],\nonumber\\
&F_{\mathrm{id}}\left[\widetilde{\rho}\right]=
\int d{\bf r}\frac{\widetilde{\rho}({\bf r})}{\gamma^2}\left\{
\ln\frac{\widetilde{\rho}({\bf r})}{\gamma^2}-1\right\},
\label{rescaled ry}
\end{flalign}
where $\widetilde{\Delta\rho}\equiv\widetilde{\rho}-\widetilde{n}$ and $n\equiv\widetilde{n}/\gamma^2$. Comparison between the rescaled functionals, eqs. (\ref{rescaled delta f}) and (\ref{rescaled ry}), suggests that the $\widetilde{\rho}$-dependent terms can be treated perturbatively at strong coupling ($\gamma\gg 1$).

\subsection{The non-equilibrium chemical potential of non-interacting spheres at strong coupling}
Going back to eq. (\ref{f rhophi def}), we devlop the $1/\gamma$ expansion method.
It is found from eq. (\ref{prescribed density}) that
\begin{flalign}
e^{\mu+i\phi({\bf r}_i)-\psi_{\mathrm{dft}}({\bf r}_i)}&=
\rho({\bf r})e^{\Delta\psi({\bf r})+i\phi({\bf r})},\nonumber\\
\Delta\psi({\bf r})
&=-\int d{\bf r}'c({\bf r}-{\bf r}')\Delta\rho({\bf r}').
\label{shifted ref potential}
\end{flalign}
Also, we shift the fluctuating-potential field from $\phi$ to $\varphi$ such that
\begin{equation}
\frac{w(0)}{2}+i\varphi({\bf r})=\Delta\psi({\bf r})+i\phi({\bf r}),
\label{varphi def}
\end{equation}
whose rescaled form is
\begin{equation}
\frac{\gamma^2\widetilde{w}(0)}{2}+i\gamma\widetilde{\varphi}({\bf r})
=\widetilde{\Delta\psi}({\bf r})+i\gamma\widetilde{\phi}({\bf r}),
\label{varphi def2}
\end{equation}
due to the relations (\ref{phi scaling}) and (\ref{chrho scaling}).
Substituting eq. (\ref{varphi def2}) into eq. (\ref{shifted ref potential}), we have the rescaled form,
\begin{flalign}
e^{\mu+i\phi({\bf r}_i)-\psi_{\mathrm{dft}}({\bf r}_i)}
=\rho({\bf r})e^{\frac{w(0)}{2}+i\varphi({\bf r})}=\frac{\widetilde{\rho}({\bf r})}{\gamma}e^{i\gamma\widetilde{\varphi}({\bf r})},
\label{shifted ref potential2}
\end{flalign}
considering that $\rho({\bf r})e^{\frac{w(0)}{2}}=\gamma\rho({\bf r})=\widetilde{\rho}({\bf r})/\gamma$. Moreover, eqs. (\ref{prescribed density}), (\ref{varphi def}) and (\ref{shifted ref potential2}) are arranged to give
\begin{flalign}
&-\int d{\bf r}\rho({\bf r})\left\{\psi_{\mathrm{dft}}({\bf r})-i\phi({\bf r})\right\}\nonumber\\
&\qquad=\int d{\bf r}\rho({\bf r})\left\{\ln\rho({\bf r})+\Delta\psi({\bf r})+i\phi({\bf r})-\mu\right\}\nonumber\\
&\qquad=\int d{\bf r}\rho({\bf r})\left\{\ln\rho({\bf r})+\frac{w(0)}{2}+i\varphi({\bf r})-\mu\right\}\nonumber\\
&\qquad\equiv F_{0}[\rho,\varphi]-\int d{\bf r}\rho({\bf r})\mu.
\label{shift ext}
\end{flalign}
Combining eqs. (\ref{shifted ref potential}) to (\ref{shift ext}), eq. (\ref{f rhophi def}) reads
\begin{flalign}
&e^{-F\left[\rho,\phi=\varphi+i\Delta\psi\right]+\int d{\bf r}\rho({\bf r})\mu}\nonumber\\
&\qquad=e^{-F_{0}[\rho,\varphi]+\int d{\bf r}\rho({\bf r})\mu}\nonumber\\
&\qquad\qquad\times\mathrm{Tr}\prod_i\frac{\widetilde{\rho}({\bf r}_i)}{\gamma}e^{i\gamma\widetilde{\varphi}({\bf r}_i)}\prod_{i,j}e^{-v({\bf r}_i-{\bf r}_j)},
\label{mayer1 f rhophi}
\end{flalign}
which is the functional to be evaluated using the $1/\gamma$ expansion method at strong coupling, $\gamma\gg 1$.

Let us see the non-equilibrium chemical potential $\lambda[\rho]$ in a reference system of non-interacting spheres, prior to formulating the strong-coupling approximation of eq. (\ref{mayer1 f rhophi}). In the absence of the interaction potential $v({\bf r}_i-{\bf r}_j)$, eq. (\ref{mayer1 f rhophi}) is exactly reduced to the ideal free-energy functional for non-interacting system:
\begin{flalign}
&F_{\mathrm{non}}[\rho,\varphi]\nonumber\\
&=F_{0}[\rho,\varphi]-\int d{\bf r}\frac{\widetilde{\rho}({\bf r})}{\gamma}e^{i\gamma\widetilde{\varphi}({\bf r})}\nonumber\\
&=\int d{\bf r}\widetilde{\rho}({\bf r})\left\{
\frac{1}{\gamma^2}\ln\frac{\widetilde{\rho}({\bf r})}{\gamma^2}
+\frac{\widetilde{w}(0)}{2}+\frac{i\widetilde{\varphi}({\bf r})}{\gamma}
-\frac{e^{\frac{\gamma^2\widetilde{w}(0)}{2}+i\gamma\widetilde{\varphi}({\bf r})}}{\gamma^2}
\right\}\nonumber\\
&= F_{0}[\widetilde{\rho},\widetilde{\varphi}]
-\frac{1}{\gamma^2}\int d{\bf r}\widetilde{\rho}({\bf r})
e^{\frac{\gamma^2\widetilde{w}(0)}{2}+i\gamma\widetilde{\varphi}({\bf r})}.
\label{ideal f rhophi}
\end{flalign}
It follows that
\begin{flalign}
&\lambda_{\mathrm{non}}[\widetilde{\rho},\widetilde{\varphi}]\equiv\frac{\delta
F_{\mathrm{non}}[\rho,\varphi]}{\delta\rho}\nonumber\\
&=\gamma^2\frac{\delta
F_{0}[\widetilde{\rho},\widetilde{\varphi}]}{\delta\widetilde{\rho}}-
\frac{\delta}{\delta\widetilde{\rho}({\bf r})}\left\{\int d{\bf r}
\widetilde{\rho}({\bf r})
e^{\frac{\gamma^2\widetilde{w}(0)}{2}+i\gamma\widetilde{\varphi}({\bf r})}
\right\}\nonumber\\
&=\ln\frac{\widetilde{\rho}({\bf r})}{\gamma^2}+1+\frac{\gamma^2\widetilde{w}(0)}{2}+i\gamma\widetilde{\varphi}({\bf r})\nonumber\\
&\hphantom{=\gamma^2\frac{\delta
F_{0}[\widetilde{\rho},\widetilde{\varphi}]}{\delta\widetilde{\rho}}}
-e^{\frac{\gamma^2\widetilde{w}(0)}{2}+i\gamma\widetilde{\varphi}({\bf r})}-\widetilde{\rho}({\bf r})\frac{\delta e^{\frac{\gamma^2\widetilde{w}(0)}{2}+i\gamma\widetilde{\varphi}({\bf r})}}{\delta\widetilde{\rho}({\bf r})}.
\label{ideal diff f rhophi}
\end{flalign}
We need to perform the average of $\lambda_{\mathrm{non}}[\widetilde{\rho},\widetilde{\varphi}]$ over the $\phi$--field based on the original definition in addition to the relation (\ref{varphi def2}). In the strong-coupling approximation, we obtain from eq. (\ref{ideal diff f rhophi})
\begin{flalign}
\lambda_{\mathrm{non}}[\rho]=\overline{\lambda_{\mathrm{non}}[\widetilde{\rho},\widetilde{\varphi}]}
=\ln\rho({\bf r})+\frac{w(0)}{2},
\label{ideal chem potential}
\end{flalign}
which corresponds to the non-equilibrium chemical potential $\lambda_{\mathrm{non}}[\rho]$ of non-interacting spheres; see Appendix C2 for the detailed derivation of eq. (\ref{ideal chem potential}).
Equation (\ref{ideal chem potential}) implies that
\begin{flalign}
F_{\mathrm{non}}[\rho]
=F_{\mathrm{id}}[\rho]+\int d{\bf r}\,\rho({\bf r})\frac{w(0)}{2}.
\label{non free}
\end{flalign}
 \subsection{Connecting the $1/\gamma$ expansion with the virial-type expansion: derivation scheme of eqs. (\ref{lambda result}) and (\ref{c result})}
In the strong-coupling approximation, long-range correlations of the shifted fluctuating potential $\varphi({\bf r})$ is maintained:
\begin{flalign}
\overline{\varphi({\bf r})\varphi({\bf r}')}&=w({\bf r}-{\bf r}'),
\label{conclusion varphivarphi}
\end{flalign}
as well as the relation (\ref{f second w}) for the fluctuating $\phi$--potential (see the derivation of eq. (\ref{appendix varphivarphi}) in Appendix C1).
Equation (\ref{conclusion varphivarphi}) implies that
\begin{equation}
\left|\frac{\widetilde{\rho}({\bf r}_i)}{\gamma}e^{i\gamma\widetilde{\varphi}({\bf r}_i)}\right|
=\frac{\widetilde{\rho}({\bf r}_i)}{\gamma}.
\label{absolute}
\end{equation}
Hence, the $1/\gamma$ expansion becomes equivalent to the following density expansion (or the fugacity expnasion \cite{netz}):
\begin{flalign}
&\mathrm{Tr}\prod_i\frac{\widetilde{\rho}({\bf r}_i)}{\gamma}e^{i\gamma\widetilde{\varphi}({\bf r}_i)}\prod_{i,j}e^{-v({\bf r}_i-{\bf r}_j)}\nonumber\\
&=\sum_{N=0}^{\infty}\frac{1}{N!}\int d{\bf r}_1\cdots\int d{\bf r}_N\prod_{i=1}^N\frac{\widetilde{\rho}({\bf r}_i)}{\gamma}\,e^{i\gamma\widetilde{\varphi}({\bf r}_i)}\prod_{i,j}e^{-v({\bf r}_i-{\bf r}_j)}\nonumber\\
&\approx
1+\frac{1}{\gamma}\left\{
\int d{\bf r}\widetilde{\rho}({\bf r})\,e^{i\gamma\widetilde{\varphi}({\bf r})}
\right.\nonumber\\
&\left.
+\frac{1}{2\gamma}\iint d{\bf r}_1d{\bf r}_2\widetilde{\rho}({\bf r}_1)\widetilde{\rho}({\bf r}_2)\,e^{-v({\bf r}_1-{\bf r}_2)+i\gamma\widetilde{\varphi}({\bf r}_1)+i\gamma\widetilde{\varphi}({\bf r}_2)}
\right\}
\nonumber\\
&\equiv
1+\frac{1}{\gamma}U[\widetilde{\rho},\widetilde{\varphi}],
\label{density expansion}
\end{flalign}
where
\begin{flalign}
&U[\widetilde{\rho},\widetilde{\varphi}]\nonumber\\
&=\int d{\bf r}\widetilde{\rho}({\bf r})\,e^{i\gamma\int d{\bf s}\,\widetilde{\varphi}({\bf s})\widehat{\rho}_1({\bf s})}\nonumber\\
&+\frac{1}{2\gamma}\iint d{\bf r}_1d{\bf r}_2\widetilde{\rho}({\bf r}_1)\widetilde{\rho}({\bf r}_2)\,e^{-v({\bf r}_1-{\bf r}_2)+i\gamma\int d{\bf s}\,\widetilde{\varphi}({\bf s})\widehat{\rho}_2({\bf s})}.
\label{u form}
\end{flalign}
In eq. (\ref{u form}), we have introduced instantaneous one-and two-particle densities, $\widehat{\rho}_1({\bf s})=\delta({\bf s}-{\bf r})$ and $\widehat{\rho}_2({\bf s})=\sum_{i=1}^2\delta({\bf r}-{\bf r}_i)$, for making a distinction between the first and second terms on the right hand side of eq. (\ref{u form}).

Combination of eqs. (\ref{mayer1 f rhophi}), (\ref{ideal f rhophi}) and (\ref{density expansion}) provides
\begin{flalign}
F[\widetilde{\rho},\widetilde{\varphi}]
=F_{0}[\widetilde{\rho},\widetilde{\varphi}]
-\ln
\left(1+\frac{1}{\gamma}U[\widetilde{\rho},\widetilde{\varphi}]\right),
\label{difference sc}
\end{flalign}
and we define the non-equilibrium chemical potential difference $\Delta\lambda[\rho]$ due to the addition of the interaction potential $v({\bf r})$ as follows:
\begin{flalign}
\Delta\lambda[\rho]
=\gamma^2\overline{\frac{\delta
F[\widetilde{\rho},\widetilde{\varphi}]}
{\delta\widetilde{\rho}({\bf r})}}
-\lambda_{\mathrm{non}}[\rho].
\label{delta lambda2}
\end{flalign}
It follows from eqs. (\ref{ideal f rhophi}), (\ref{ideal chem potential}), (\ref{conclusion varphivarphi}), (\ref{u form}) and (\ref{difference sc}) that the strong-coupling approximation of eq. (\ref{delta lambda2}) leads to
\begin{flalign}
&\Delta\lambda[\rho]\nonumber\\
&=1-\overline{\frac{\gamma\delta
U[\widetilde{\rho},\widetilde{\varphi}]/\delta\widetilde{\rho}({\bf r})}{1+U[\widetilde{\rho},\widetilde{\varphi}]/\gamma}}\nonumber\\
&=
1-\overline{\frac{\gamma\delta
U[\widetilde{\rho},\widetilde{\varphi}]}{\delta\widetilde{\rho}({\bf r})}}
+\overline{\frac{\delta
U[\widetilde{\rho},\widetilde{\varphi}]}{\delta\widetilde{\rho}({\bf r})}
U[\widetilde{\rho},\widetilde{\varphi}]}+\mathcal{O}[\gamma^{-1}]
\nonumber\\
&=
1-\gamma\overline{
e^{i\gamma\int d{\bf s}\,\widetilde{\varphi}({\bf s})\widehat{\rho}_1({\bf s})}}\nonumber\\
&\qquad-\int d{\bf r}_2\,\widetilde{\rho}({\bf r}_2)e^{-v({\bf r}_1-{\bf r}_2)}\overline{
e^{i\gamma\int d{\bf s}\,\widetilde{\varphi}({\bf s})\widehat{\rho}_2({\bf s})}}\nonumber\\
&\qquad+\int d{\bf r}'\widetilde{\rho}({\bf r}')\,\overline{e^{i\gamma\int d{\bf s}\,\widetilde{\varphi}({\bf s})\left\{\widehat{\rho}_1({\bf s})+\widehat{\rho'}_1({\bf s})\right\}}}+\mathcal{O}[\gamma^{-1}]
\nonumber\\
&=\int d{\bf r}'\left\{1-e^{-v({\bf r}-{\bf r}')-w({\bf r}-{\bf r}')}\right\}\rho({\bf r}')+\mathcal{O}\left[\gamma^{-2}\right],
\label{derivative difference sc}
\end{flalign}
where it is noted that the instantaneous one-particle densities, $\widehat{\rho}_1({\bf s})=\delta({\bf s}-{\bf r})$ and $\widehat{\rho'}_1({\bf s})=\delta({\bf s}-{\bf r}')$, are unable to coexist at the same time by definition;
see Appendix C for the detailed and more precise discussions regarding the derivation of eq. (\ref{derivative difference sc}). We obtain from eqs. (\ref{ideal chem potential}) and (\ref{derivative difference sc})
\begin{flalign}
&\lambda[\rho]=\lambda_{\mathrm{non}}[\rho]+\Delta\lambda[\rho]\nonumber\\
&=\ln\rho({\bf r})+\frac{w(0)}{2}+\int d{\bf r}'\left\{1-e^{-v({\bf r}-{\bf r}')-w({\bf r}-{\bf r}')}\right\}\rho({\bf r}').
\label{shifted av}
\end{flalign}
The main result given by eqs. (\ref{lambda result}) and (\ref{c result}) is thus verified.

\section{Discussions}
In this section, we aim to gain insight into the short-range cutoff of the metastable DCF $c^*({\bf r})$ from dynamic aspects. We consider a fluctuating displacement field ${\bf u}({\bf r},t)$ which is related to a density difference, $\nu({\bf r},t)=\rho({\bf r},t)-\rho^*({\bf r})$. Since the fluctuating density field $\nu({\bf r},t)$ obeys the linearized Dean-Kawasaki equation of the stochastic DFT \cite{witt,podgornik,demery,kruger,frusawa1,frusawa2,goles}, the short-range dynamics of ${\bf u}({\bf r},t)$ can be inferred from the $\nu$--field dynamics. First, we will see that the short-range cutoff of the metastable DCF implies the disappearance of interaction-induced restoring force against the fluctuating density field $\nu({\bf r},t)$ (Sec. VIIIA). Next, connection of the short-range softening with anharmonic soft modes will be discussed in terms of the ${\bf u}$--field dynamics (Sec. VIIIB). Last, we summarize the results presented so far using Table 2 (Sec. VIIIC).

\subsection{Dynamic implication for the short-range cutoff of the metastable DCF $c^*({\bf r})$}
Expanding the non-equilibrium excess chemical potential $\lambda_{\mathrm{ex}}[\rho]$ around $\rho^*({\bf r})$, the Dean-Kawasaki equation (\ref{dean kawasaki}) becomes
\begin{flalign}
&\frac{\partial \rho({\bf r},t)}{\partial
t}=\frac{\partial\{\nu({\bf r},t)+\rho^*({\bf r})\}}{\partial
t}\nonumber\\
&=\nabla\cdot\mathcal{D}\rho\nabla
\left\{\lambda_{\mathrm{ex}}^*+\int d{\bf r}'
\left.\frac{\delta\lambda_{\mathrm{ex}}[\rho]}{\delta\rho({\bf r}')}\right|_{\rho=\rho^*}\nu({\bf r}',t)
\right\}+\zeta[\rho,\overrightarrow{\eta}]\nonumber\\
&=\nabla\cdot\mathcal{D}\rho\nabla\int d{\bf r}'
\left.\frac{\delta\lambda_{\mathrm{ex}}[\rho]}{\delta\rho({\bf r}')}\right|_{\rho=\rho^*}\nu({\bf r}',t)
+\zeta[\rho,\overrightarrow{\eta}],
\label{prelinear dk}\\
&\left.\frac{\delta\lambda_{\mathrm{ex}}[\rho]}{\delta\rho({\bf r}')}\right|_{\rho=\rho^*}
=\frac{\delta({\bf r}-{\bf r}')}{\rho^*({\bf r})}-c^*({\bf r}-{\bf r}')
-\left.\frac{\delta\lambda_N}{\delta\rho({\bf r}')}\right|_{\rho=\rho^*}\nonumber\\
&\hphantom{\left.\frac{\delta\lambda_{\mathrm{ex}}[\rho]}{\delta\rho({\bf r}')}\right|_{\rho=\rho^*}}=\frac{\delta({\bf r}-{\bf r}')}{\rho^*({\bf r})}-c^*({\bf r}-{\bf r}'),
\label{metalambda dev}
\end{flalign}
where $|\nabla\lambda_{\mathrm{ex}}^*|=0$ has been used in eq. (\ref{prelinear dk}). Equation (\ref{prelinear dk}) with eq. (\ref{metalambda dev}) leads to the linearized Dean-Kawasaki equation as follows:
\begin{flalign}
&\frac{\partial \nu({\bf r},t)}{\partial
t}=\mathcal{D}\nabla^2\nu({\bf r},t)\nonumber\\
&-\nabla\cdot \mathcal{D}\rho^*\int d{\bf r}'\nabla c^*({\bf r}-{\bf r}')\nu({\bf r}',t)
+\sqrt{2}\zeta[\rho^*,\overrightarrow{\eta}],\nonumber\\
\label{linear dk}
\end{flalign}
due to the manipulation of noise term \cite{frusawa1}.
Equation (\ref{linear dk}) represents the overdamped dynamics of the fluctuating density field $\nu({\bf r},t)$ around a metastable non-hyperuniform state.

Equation (\ref{linear dk}) indicates that the interaction-induced restoring force against the density deviation $\nu({\bf r},t)$ is given by the sum of $-\nabla c^*({\bf r}-{\bf r}')\nu({\bf r}',t)$. Focusing on the short-range contribution to this force, we find that microscopic environments in the hyperuniform and non-hyperuniform states are quite different from each other. While the scaling behavior (\ref{small hyper dcf}) in a hyperuniform state predicts the divergence of $|\nabla c({\bf r}-{\bf r}')|\rightarrow\infty$ in the limit of $|{\bf r}-{\bf r}'|\rightarrow 0$, the short-range cutoff of the non-hyperuniform DCF $c^*({\bf r})$ creates the opposite situation on the particle scale:
\begin{flalign}
|\nabla c^*({\bf r}-{\bf r}')|\approx 0\quad(|{\bf r}-{\bf r}'|<\sigma),
\label{no force}
\end{flalign}
as seen from Fig. 1. The above relation implies that there is no interaction-induced restoring force against the $\nu$--field in non-hyperuniform states at the particle-scale while preserving the long-range contribution, $-\nabla c^*({\bf r}-{\bf r}')\nu({\bf r}',t)=\nabla (\beta/|{\bf r}-{\bf r}'|^2)\nu({\bf r}',t)$, for $|{\bf r}-{\bf r}'|\gg\sigma$.

\subsection{Microscopic mechanism behind the appearance of eq. (\ref{no force})}
Equation (\ref{linear dk}) for the overdamped Brownian dynamics is insufficient for a descriptor of vibrational modes due to the absence of the inertia term, and yet eq. (\ref{no force}) suggests the emergence of dynamic softening in non-hyperuniform systems. We can learn the microscopic mechanism of soft modes from previous studies on quasicontacts of a contact network, a skeleton of jammed matter \cite{hecke rev,liu rev,wyart rev,lub2017,lub2018,behringer rev,makse}. The previous findings could provide an intuitive understanding of the virial-type expansion at high density as will be seen below.

For isostatic and hyperuniform systems, the packing geometry uniquely defines the contact forces as well as the spatial network structures including void distributions \cite{t revjcp,t physrep,t disk,t void1,t void2,t rattler,t marginal}.
Previous studies have shown that the isostatic and hyperuniform state disappears upon relaxing the strict constraints on the size-and spatial-distributions of voids slightly away from jamming \cite{t revjcp,t physrep,t void1,t void2,t marginal}.
The relative abundance of non-isostatic contacts provides quasicontacts that carry weak forces, thereby creating local excitations with little restoring forces \cite{t marginal,liu rev,wyart rev,lub2017,lub2018,makse}.

The possible particles forming the quasicontacts include rattlers and/or bucklers \cite{t marginal,liu rev,wyart rev,lub2017,lub2018,makse,b char2015,b franz2015,b char2016,b hexner,b franz2020,b char2020,b liu}.
It has been found, for instance, that the bucklers in the $d$--dimensional space, having $d+1$ contacts as part of the contact network, are likely to be buckled to generate quasi-localized soft modes observed in the lowest-frequency regime \cite{liu rev,wyart rev,lub2017,lub2018,makse,b char2015,b franz2015,b char2016,b hexner,b franz2020,b char2020,b liu}. The anomalous vibrational modes have been shown to exhibit strong anharmonicities that are accompanied by intermittent rearrangements of particles as follows: opening a weak contact of a buckler (i.e., buckling) yields a disordered core of a few particle scale with a power-law decay of displacements which are coupled to the elastic background of the contact network \cite{liu rev,wyart rev,lub2017,lub2018,ikeda2017,ikeda2018,ikeda2020 prr,ikeda2020 jcp,ikeda2020 sm,lerner2013,lerner2016,lerner2020 prl,lerner2020 pre,lerner2020 pnas,urbani2021,berthier2016,szamel nc,berthier2019,tan sm2021,makse,b char2015,b franz2015,b char2016,b hexner,b franz2020,b char2020,b liu}.

It is not the center of our concern whether or not the rattlers and/or bucklers significantly contribute to the quasicontacts to degrade the hyperuniformity.
It is, however, illuminating to interpret eqs. (\ref{linear dk}) and (\ref{no force}) in terms of the quasi-localized soft modes.

Then, let ${\bf u}({\bf r},t)$ be a fluctuating displacement field induced by a fluctuating density field $\nu({\bf r},t)$. In the first approximation, we have \cite{s gab}
\begin{flalign}
\nu({\bf r},t)=-\nabla\cdot\left\{
\rho^*({\bf r}){\bf u}({\bf r},t)
\right\}.
\label{displacement}
\end{flalign}
Combining eqs. (\ref{main approx}), (\ref{fig1 dcf}),
(\ref{linear dk}), (\ref{no force}) and (\ref{displacement}), we
can verify that the displacement field ${\bf u}({\bf r},t)$
shares common features with that of the quasi-localized soft
modes as follows:
\begin{itemize}
\item The interaction-induced restoring force of ${\bf u}({\bf r},t)$ is long-ranged in correspondence with recent simulations \cite{s chak prl,s tanaka,lerner2020  jcp,shimada stress} because of the power-law decay of the metastable DCF $c^*({\bf r})$ represented by eqs. (\ref{main approx}) and (\ref{fig1 dcf}). 
\item Equations (\ref{linear dk}) to (\ref{displacement}) imply the short-range anharmonicity of ${\bf u}({\bf r},t)$ at the particle scale.
\end{itemize}
This connection of our theoretical results (particularly, eq. (\ref{no force})) with the quasi-localized soft modes suggests that the virial-type expansion in a glassy state represents particle-particle interactions occurring due to the intermittent particle rearrangements.

\begin{table*}
\caption{Comparison between the theoretical approaches and results. We follow the notation of the type names given in Table 1.}
\footnotesize
\begin{tabular}{cc|c|cc|ccc}
\multicolumn{2}{c|}{Theory}&
\hspace{2pt}State description$\>$&
\multicolumn{2}{|c|}{Statics}&\multicolumn{3}{c}{Dynamics}\\[2pt]
\hline
&&&&&&&\\
DFT type$\>$&
\hspace{1pt}Approximation$\>$&
\hspace{1pt}Type$\>$&
\hspace{1pt}$\mu$ or $\lambda^*\quad$&
\hspace{6pt}DCF$\>$&
\hspace{1pt}Equation$\>$&
\hspace{1pt}Short-range$\>$&
\hspace{1pt}Long-range\\[8pt]
\hline\hline
&&&&&&&\\
Equilibrium$\>$&
\hspace{1pt}Ramakrishnan-Yussouff$\>$&
\hspace{1pt}H2$\>$&
\hspace{1pt}Eq. (\ref{dft chem2})$\quad$&
\hspace{6pt}Eq. (\ref{h2 dcf}) or (\ref{fig1 dcf})$\>$&
\hspace{1pt}Eq. (\ref{dean kawasaki2})$\>$&
\hspace{1pt}Frozen$\>$&
\hspace{1pt}Correlated\\[10pt]
Stochastic$\>$&
\hspace{1pt}Gaussian$\>$&
\hspace{1pt}H2$\>$&
\hspace{1pt}Eq. (\ref{lambda gaussian result})$\quad$&
\hspace{6pt}Eq. (\ref{h2 dcf}) or (\ref{fig1 dcf})$\>$&
\hspace{1pt}Eq. (\ref{dean kawasaki}) or (\ref{linear dk})$\>$&
\hspace{1pt}Frozen$\>$&
\hspace{1pt}Correlated\\[10pt]
\hline
&&&&&&&\\
\underline{\bf Stochastic}$\>$&
\hspace{1pt}\underline{\bf Strong-coupling}$\>$&
\hspace{1pt}\fbox{\bf N1}$\>$&
\hspace{1pt}Eq. (\ref{lambda result})$\quad$&
\hspace{6pt}Eq. (\ref{c result}) or (\ref{main approx})$\>$&
\hspace{1pt}Eq. (\ref{dean kawasaki}) or (\ref{linear dk})$\>$&
\hspace{1pt}\underline{\bf Soft}$\>$&
\hspace{1pt}Correlated\\[8pt]
\end{tabular}
\end{table*}

\subsection{Summarizing the results in comparison with other treatments}
The differences in the free-energy density functionals between the equilibrium and stochastic DFTs are summarized as follows: 
\begin{itemize} 
\item[(i)] The density functional $\mathcal{A}[\rho]$ appearing in the metastability equation (\ref{metastable}) represents the free-energy functional of a given density distribution $\rho({\bf r})$, instead of the equilibrium free-energy functional $F[\rho,0]$. It is a clear advantage over the equilibrium DFT that the stochastic DFT can make use of a field-theoretic formulation in obtaining $\mathcal{A}[\rho]$.
\item[(ii)] The metastability equation (\ref{metastable}) states that the functional derivative $\delta\mathcal{A}[\rho]/\delta\rho$ should yield a spatially constant $\lambda_{\mathrm{ex}}^*$, which has been referred to as the metastable excess chemical potential. The sum of $\lambda_{\mathrm{ex}}^*$ and the Lagrange multiplier $\lambda_N$ corresponds to the metastable chemical potential and is reduced to the equilibrium chemical potential (i.e., $\lambda_{\mathrm{ex}}^*+\lambda_N=\mu$) when $\lambda_{\mathrm{ex}}^*=0$ and $\lambda_N=\mu$ in equilibrium; see also the discussion after eq. (\ref{eq excess}).
\item[(iii)] As found from eqs. (\ref{dft metastable}), (\ref{s eq approx}) and (\ref{fig1 dcf}), the input of the hyperuniform DCF allows the equilibrium DFT to predict the hyperuniformity of a metastable state, without the knowledge on the reference density distribution in an amorphous state.
\end{itemize}

These characteristics of the stochastic DFT enable us to evaluate the extent to which fluctuations around the metastable density $\rho^*$ affect the metastable chemical potential $\lambda^*$. Actually, we have demonstrated that the stochastic DFT is relevant to determine metastable states around a hyperuniform state. The stochastic DFT provides an analytical form of the metastable DCF that has a short-range cutoff inside the sphere while retaining the long-range power-law behavior. We should keep in mind that the long-range hyperuniform behavior is preserved because the Gaussian weight, $e^{-\Delta F_{\mathrm{dft}}}$, for the virial-type expansion premises that non-hyperuniform states considered are located near a hyperuniform state. As confirmed in Sec. VI, the obtained DCF yields the zero-wavevector structure factor in quantitative agreement with previous simulation results \cite{t revjcp,t physrep,ozawa,chieco,silbert,olsson,ikeda non1,ikeda non2} of degraded hyperuniformity.

Moreover, both Fig. 3 and Table 2 summarize the results by comparing the following theoretical approaches discussed so far: the equilibrium DFT using the Ramakrishnan-Yussouff free-energy functional \cite{ry}, the stochastic DFT in the Gaussian approximation (see Sec. V), and the stochastic DFT in the strong-coupling approximation (see Sec. VII). 
\begin{figure}[hbtp]
\begin{center}
	\includegraphics[
	width=9cm
]{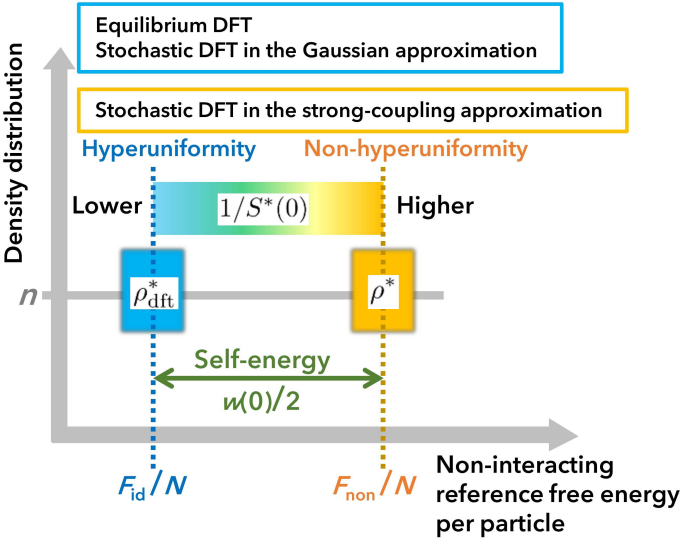}
\end{center}
\caption{
A schematic comparison of theoretical approaches presented in this study. The spatially uniform density is identically $n$ as shown on the vertical axis, and the hyperuniformity is incorporated into the equilibrium DFT by inputting the hyperuniform DCF $c({\bf r})$ given by eq. (\ref{fig1 dcf}); nevertheless, we have hyperuniform and non-hyperuniform treatments colored blue and orange, respectively. The different results are due to distinct values of non-interacting reference free-energy functionals (i.e., $F_{\mathrm{id}}[\rho]$ given by eq. (\ref{ry functional}) and $F_{\mathrm{non}}[\rho]$ given by eq. (\ref{non free})), which is represented by the horizontal axis.
}
\end{figure}

The vertical axis in Fig. 3 shows that the density distribution considered has the same density $n$ on average. The difference is attributed to the inhomogeneous distributions around $n$: the hyperuniform density distribution $\rho^*_{\mathrm{dft}}({\bf r})$, which is colored blue, satisfies eq. (\ref{hyper result}) for the inverse of the zero-wavevector structure factor $1/S^*(0)$, whereas the non-hyperuniform range of density distribution $\rho^*({\bf r})$, which is colored orange, satisfies eq. (\ref{nonhyper result}). As summarized in Fig. 3, the equilibrium DFT and the stochastic DFT in the Gaussian approximation provides the hyperunifomity, whereas the stochastic DFT in the strong-coupling approximation the non-hyperuniformity.

Meanwhile, the transverse axis of Fig. 3 shows that the above two types of theoretical approaches take distinct reference free-energy functionals, as found from comparing $F_{\mathrm{id}}$ and $F_{\mathrm{non}}$ given by eqs. (\ref{ry functional}) and (\ref{non free}), respectively.
In the hyperuniform theories, on the one hand, the ideal free-energy functional $F_{\mathrm{id}}/N$ per particle is the reference functional for the evaluation of interaction energy (see eq. (\ref{ry functional}), following the conventional treatment of the equilibrium DFT \cite{ry,evans,singh,likos,lut1}, where $N$ denotes the total number of spheres as before. On the other hand, as a reference functional of the stochastic DFT in the strong-coupling approximation, we used the free-energy functional $F_{\mathrm{non}}/N$ of a non-interacting system per particle that is larger than the ideal one $F_{\mathrm{non}}/N$ by the self-energy $w(0)/2$.
It can be stated that a perturbation field theory method becomes more relevant to the evaluation of intermittent fluctuations, due to the increase in the reference free energy.

Table 2 presents more detailed classifications of the hyperuniform and non-hyperuniform theories. There are two types of classifications for the above three treatments.
One classification is based on the DFT type of whether the dynamical DFT relies on the deterministic equation (\ref{dean kawasaki2}) or the stochastic equation (\ref{dean kawasaki}). The former approach represented by eq. (\ref{dean kawasaki2}) is equivalent to the equilibrium DFT as clarified at the beginning of Sec. IIID, whereas the latter equation (\ref{dean kawasaki}) forms the basis of the last two stochastic approaches where the additional contribution $\Delta F[\rho,\phi]$ to the intrinsic Helmholtz free energy $F[\rho,0]$ is to be considered.
The other aspect of theoretical classification concerns the predictability of non-hyperuniformity especially when the hyperuniformity is incorporated into the DCF (i.e., eq. (\ref{h2 dcf}) or (\ref{fig1 dcf})) of the equilibrium DFT as input. The type specification column in Table 2 indicates that the stochastic DFT in the Gaussian approximation falls into the same category (type H2 defined in Table 1) of the equilibrium DFT in this light.

As confirmed from Table 2, the use of density-expansion method at strong coupling is indispensable to convert the hyperuniform structure factor at zero wavevector into the non-hyperuniform one satisfying the simulation results given by eq. (\ref{nonhyper result}). The outstanding feature of the non-hyperuniform DCF $c^*({\bf r})$ is the short-range cutoff, thereby predicting the absence of interaction-induced restoring force for the short-range dynamics (i.e., eq. (\ref{no force})).

\section{Concluding remarks}
For comparison purposes, let us go back to the previous study \cite{t kim} where the degradation of perfect hyperuniformity in crystals, quasicrystals, and disordered packings has been demonstrated both theoretically and numerically, using the three scenarios of the imperfections (see Sec. II for the list of the scenarios). The second scenario (ii), which has been our concern, attributes the violation of hyperuniformity to the stochastic occurrence of spatially correlated displacements.

Combination of eq. (\ref{conclusion varphivarphi}) and the dynamical discussions in Sec. VIII suggests that the above second scenario for the degradation of hyperuniformity is similar to the underlying physics described by the averaged virial-type interaction term, the third term on the rhs of eq. (\ref{lambda result}), under the long-range-correlated $\varphi$--field; for it seems plausible that the long-range-correlated potential field arises from the elastic nature of the contact network.  From the discussions, we infer that the virial-type degradation of hyperuniformity reflects intermittent rearrangements of particles and is more likely to be found in the collectively jammed packings allowing for the shear deformations as mentioned before, rather than in the strictly jammed ones \cite{t revjcp,t physrep, t disk}. In other words, the type H2 defined in Table 1 corresponds to the collectively jammed packings, whereas the type H1 to the strictly jammed packings.

It remains to be seen whether the present formulation can be extended to address non-hyperuniform behaviors of other measures, than the density-density structure factor, which are obtained from various physical quantities including the local number variance for a window \cite{t preprint} and the contact number fluctuations \cite{two div,parisi2021}.
We also envision that advancing the stochastic DFT \cite{witt,dean,seifert,kim,jac,das2015,podgornik,demery,kruger,lut science,frusawa1,frusawa2,goles} at strong coupling will pave the way for a realistic description of the quasi-localized soft modes induced by intermittent rearrangements of particles such as bucklers \cite{liu rev,wyart rev,lub2017,lub2018,makse,b char2015,b franz2015,b char2016,b hexner,b franz2020,b char2020,b liu}.

\appendix
\section{Details on the constrained free-energy functional $\mathcal{A}[\rho]$}
\subsection{Verification of eq. (\ref{pst})}
The distribution functional $P[\rho,t]$ defined by eq. (\ref{p def}) satisfies the Fokker-Planck equation as follows \cite{witt,frusawa0}:
\begin{eqnarray}
\frac{\partial P[\rho,t]}{\partial t}=-\int d{\bf r}
\frac{\delta}{\delta\rho}\nabla\cdot
\mathcal{D}\rho\,\nabla
\left[
\frac{\delta}{\delta\rho}+\frac{\delta
\mathcal{A}[\rho]}{\delta\rho}
\right]
 P[\rho,t],\nonumber\\
\label{appendix fokker}
\end{eqnarray}
from which we find that eq. (\ref{pst}) satisfies the stationary condition $\partial P_{\mathrm{st}}[\rho]/\partial t=0$. It has also been shown that eq. (\ref{appendix fokker}) is equivalent to the Dean-Kawasaki equation (\ref{dean kawasaki}) \cite{witt,frusawa0}.

 \subsection{Definition of $\mathcal{A}[\rho]$} 
In eq. (\ref{appendix fokker}) as well as in eq. (\ref{dean kawasaki}), the canonical ensemble is naturally required for the free-energy functional $\mathcal{A}[\rho]$ of a given density $\rho$ because we consider the overdamped dynamics of densely packed sphere system with the total number $N$ of spheres being fixed. Hence, $\mathcal{A}[\rho]$ is defined using the configurational integral for the canonical ensemble as follows:
\begin{flalign}
&e^{-\mathcal{A}[\rho]}
=\frac{1}{N!}\int d{\bf r}_1\cdots\int d{\bf r}_N
\prod_{i,j}e^{-v({\bf r}_i-{\bf r}_j)}\nonumber\\
&\qquad\qquad\times\prod_{{\bf r}}\delta\left[
\widehat{\rho}_N({\bf r},t)-\rho({\bf r},t)
\right].
\end{flalign}
Yet, the equilibrium DFT, a key ingredient in this study, needs to be formulated in the grand canonical system. We therefore write $\mathcal{A}[\rho]$ with the help of the contour integral over a complex variable $z=e^{\mu}$ \cite{frusawa1,frusawa0}:
\begin{flalign}
&e^{-\mathcal{A}[\rho]}\nonumber\\
&=\frac{1}{2\pi i}\oint \frac{dz}{z^{N+1}}\nonumber\\
&\qquad\times\left(\underline{
\mathrm{Tr}\prod_ie^{\mu}\prod_{i,j}e^{-v({\bf r}_i-{\bf r}_j)}\prod_{{\bf r}}\delta\left[
\widehat{\rho}_N({\bf r},t)-\rho({\bf r},t)\right]}
\right),
\label{contour}
\end{flalign}
so that the canonical ensemble may be recovered after performing the grand canonical ensemble represented by $\mathrm{Tr}\equiv\sum_{N=0}^{\infty}\frac{1}{N!}\int d{\bf r}\,_1\cdots\int d{\bf r}\,_N$.

 \subsection{Derivation of eq. (\ref{f first})} 
We evaluate the underlined term in eq. (\ref{contour}) with the help of the Fourier transform of the delta functional as follows:
\begin{flalign}
&\mathrm{Tr}\prod_{i}e^{\mu}\prod_{i,j}e^{-v({\bf r}_i-{\bf r}_j)}\prod_{{\bf r}}\delta\left[
\widehat{\rho}_N({\bf r})-\rho({\bf r})\right]\nonumber\\
&=\int D\psi\,
\,\mathrm{Tr}\,\prod_{i}e^{\mu+i\psi({\bf r}_i)}\prod_{i,j}e^{-v({\bf r}_i-{\bf r}_j)}
e^{-\int d{\bf r}\,\,i\psi({\bf r})\rho({\bf r})}\nonumber\\
&=\int D\psi\,
e^{-\Omega[-i\psi]-\int d{\bf r}\,\,i\psi({\bf r})\rho({\bf r})}.
\label{appendix given f functional}
\end{flalign}
The $\psi$--field is separated into a fluctuating potential field $\phi({\bf r})$ and the saddle-point field $i\psi_{\mathrm{dft}}({\bf r})$:
\begin{equation}
\psi({\bf r})=\phi({\bf r})+i\psi_{\mathrm{dft}}({\bf r}),
\label{appendix split}
\end{equation}
where $\psi_{\mathrm{dft}}({\bf r})$ is determined by the saddle-point equation,
\begin{align}
&\left.
\frac{\delta\left(\Omega[-i\psi]\right)}{\delta\psi({\bf r})}\right|_{\psi=i\psi_{\mathrm{dft}}}
=-i\rho({\bf r}).
\label{appendix omega psi}
\end{align}
The functional differentiation on the left hand side of eq. (\ref{appendix omega psi}) provides the density in equilibrium of the system under the external field $\psi_{\mathrm{dft}}$. Denoting the equilibrium density by $\left<\widehat{\rho}_N({\bf r})\right>_{\mathrm{eq}}$, the saddle-point equation (\ref{appendix omega psi}) implies that
\begin{equation}
\left<\widehat{\rho}_N({\bf r})\right>_{\mathrm{eq}}
=\rho({\bf r}).
\label{appendix prescribe}
\end{equation}
The above relation states that a prescribed density $\rho({\bf r})$ is equated with the equilibrium density due to the potential $\psi_{\mathrm{dft}}$ along the saddle-point field.

For later convenience, we also introduce the intrinsic Helmholtz free energy $F[\rho,0]$, the central functional of the equilibrium DFT \cite{evans,singh,likos,lut1}:
\begin{align}
&F[\rho,0]-\int d{\bf r}\rho({\bf r})\mu
\equiv\Omega[\psi_{\mathrm{dft}}]-\int d{\bf r}\,\rho({\bf r})\psi_{\mathrm{dft}}({\bf r}),
\label{appendix def f}
\end{align}
showing that the intrinsic Helmholtz free energy $F[\rho,0]$ is defined by the first Legendre transform of the grand potential $\Omega[\psi_{\mathrm{dft}}]$ with the saddle-point field $\psi_{\mathrm{dft}}({\bf r})$ being applied. Therefore, the identity (\ref{f first}) is satisfied as well as that in the equilibrium DFT.

 \subsection{Derivation of eq. (\ref{f rhophi def})} 
Combining eqs. (\ref{appendix given f functional}) and (\ref{appendix split}), we have
\begin{flalign}
&\mathrm{Tr}\prod_{i}e^{\mu}\prod_{i,j}e^{-v({\bf r}_i-{\bf r}_j)}\prod_{{\bf r}}\delta\left[
\widehat{\rho}_N({\bf r})-\rho({\bf r})\right]\nonumber\\
&=\int D\phi\,\mathrm{Tr}\prod_i e^{\mu+i\phi({\bf r}_i)-\psi_{\mathrm{dft}}({\bf r}_i)}
\prod_{i,j}e^{-v({\bf r}_i-{\bf r}_j)}\nonumber\\
&\qquad\qquad\qquad\qquad\times
e^{\int d{\bf r}\rho({\bf r})\left\{\psi_{\mathrm{dft}}({\bf r})-i\phi({\bf r})\right\}}\nonumber\\
&=\int D\phi\,e^{-\Omega[\psi_{\mathrm{dft}}-i\phi]+\int d{\bf r}\rho({\bf r})
\left\{\psi_{\mathrm{dft}}({\bf r})-i\phi({\bf r})\right\}}\nonumber\\
&=\int D\phi\,e^{-F[\rho,\phi]+\int d{\bf r}\rho({\bf r})\mu},
\label{appendix f rhophi def}
\end{flalign}
when defining 
\begin{align}
&F[\rho,\phi]-\int d{\bf r}\rho({\bf r})\mu\nonumber\\
&\equiv\Omega[\psi_{\mathrm{dft}}-i\phi]
-\int d{\bf r}\,\rho({\bf r})\left\{\psi_{\mathrm{dft}}({\bf r})-i\phi({\bf r})\right\}
\label{appendix extended f}
\end{align}
as an extension of eq. (\ref{appendix def f}). Equations (\ref{appendix f rhophi def}) and (\ref{appendix extended f}) validate eq. (\ref{f rhophi def}).

 \subsection{Derivation of eqs. (\ref{a rho def}) and (\ref{delta number})} 
It follows from eqs. (\ref{contour}) and (\ref{appendix f rhophi def}) that
\begin{flalign}
&e^{-\mathcal{A}[\rho]}\nonumber\\
&=\frac{1}{2\pi i}\oint \frac{dz}{z^{N+1}}\int
D\phi\,e^{-F[\rho,\phi]+\int d{\bf r}\rho({\bf r})\mu}\nonumber\\
&=\int D\phi\,e^{-F[\rho,\phi]}
\left(\frac{1}{2\pi i}\oint dz\,\frac{1}{z^{-\int d{\bf r}\,\rho({\bf r},t)+N+1}}\right)\nonumber\\
&=\int D\phi\,e^{-F[\rho,\phi]}\Delta[\rho],
\label{appendix a rho def}
\end{flalign}
where
\begin{eqnarray}
\Delta[\rho]&\equiv&
\frac{1}{2\pi i}\oint dz\,\frac{1}{z^{-\int d{\bf r}\,\rho({\bf r},t)+N+1}}\nonumber\\
&=&\left\{
\begin{array}{l}
1\quad(\int d{\bf r}\rho({\bf r})=N)\\
\\ 
0\quad(\int d{\bf r}\rho({\bf r})\neq N).\\
\end{array}
\right.
\label{appendix delta number}
\end{eqnarray}
Equations (\ref{appendix a rho def}) and (\ref{appendix delta number}) verify eqs. (\ref{a rho def}) and (\ref{delta number}), respectively.

 \subsection{Derivation of eq. (\ref{delta f w})} 
The difference $\Delta F_{\mathrm{dft}}[\rho,\phi]=F[\rho,\phi]-F[\rho,0]$ between eqs. (\ref{appendix def f}) and (\ref{appendix extended f}) reads
\begin{align}
\Delta
F_{\mathrm{dft}}[\rho,\phi]&=\Omega[\psi_{\mathrm{dft}}-i\phi]-\Omega[\psi_{\mathrm{dft}}]
+\int d{\bf r}\,i\rho({\bf r})\phi({\bf r}).
\label{appendix delta omega}
\end{align}
The quadratic expansion of $\Omega[\psi_{\mathrm{dft}}-i\phi]$ around the imaginary saddle-point field $i\psi_{\mathrm{dft}}$ yields
\begin{flalign}
\Delta F_{\mathrm{dft}}[\rho,\phi]
&=\frac{1}{2}\iint d{\bf r}d{\bf r}'\,
\left.
\frac{\delta^2\Omega}{\delta \psi({\bf r})\delta \psi({\bf r}')}\right|_{\psi=\psi_{\mathrm{dft}}}
\phi({\bf r})\phi({\bf r}'),
\label{hg phim}\\
&=\frac{1}{2}\iint d{\bf r}d{\bf r}'\phi({\bf r})w^{-1}({\bf r}-{\bf r}')\phi({\bf r}'),
\end{flalign}
where use has been made of the basic relation in the equilibrium DFT as follows:
\begin{flalign}
\int d{\bf r}'\frac{\delta^2 F[\rho,0]}{\delta\rho({\bf r})\delta\rho({\bf r}')}
\left.
\frac{\delta^2\Omega}{\delta \psi({\bf r}')\delta \psi({\bf r}")}
\right|_{\psi=i\psi_{\mathrm{dft}}} =\delta({\bf r}-{\bf r}"),
\label{appendix f omega}
\end{flalign}
as well as the definition (\ref{f second w}) of $w({\bf r}-{\bf r}')$.

\subsection{Derivation of eq. (\ref{phiphi transform})}
It is found from eq. (\ref{w-1 h}) that eq. (\ref{phiphi transform}) reads
\begin{flalign}
&\overline{\frac{\delta}{\delta\rho}\iint d{\bf r}d{\bf r}'w^{-1}({\bf r}-{\bf r}')
\phi({\bf r})\phi{\bf r}')}
\nonumber\\
&=\int d{\bf r}'\left[
\left\{w^{-1}({\bf r}-{\bf r}')\right\}'\overline{\phi({\bf r})\phi({\bf r}')}
\right.\nonumber\\
&\left.\qquad\qquad
+w^{-1}({\bf r}-{\bf r}')
\frac{\int D\phi\,\left\{\phi({\bf r})\phi({\bf r}')\right\}'e^{-\Delta F_{\mathrm{dft}}[\rho,\phi]}}
{\int D\phi\,e^{-\Delta F_{\mathrm{dft}}[\rho,\phi]}}\right],
\label{appendix phiphi transform}
\end{flalign}
where
\begin{flalign}
\left\{w^{-1}({\bf r}-{\bf r}')\right\}'
&\approx\delta({\bf r}-{\bf r}')+2h({\bf r}-{\bf r}')\rho({\bf r}')\nonumber\\
&=\frac{w^{-1}({\bf r}-{\bf r}')}{\rho({\bf r})}+h({\bf r}-{\bf r}')\rho({\bf r}'),
\label{dash1}
\end{flalign}
neglecting the density dependence of the total correlation function $h({\bf r}-{\bf r}')$, and eqs. (\ref{appendix phiphi transform}) and (\ref{dash1}) give a precise definition of $\left\{\phi({\bf r})\phi({\bf r}')\right\}'$ appearing in eq. (\ref{appendix phiphi transform}).

In eq. (\ref{appendix phiphi transform}), we use the following approximation:
\begin{flalign}
&\left\{\overline{\phi({\bf r})\phi({\bf r}')}\right\}'
\equiv\frac{\delta}{\delta\rho({\bf r})}\left\{
\frac{\int D\phi\,\phi({\bf r})\phi({\bf r}')e^{-\Delta F_{\mathrm{dft}}[\rho,\phi]}}
{\int D\phi\,e^{-\Delta F_{\mathrm{dft}}[\rho,\phi]}}
\right\}\nonumber\\
&=\frac{\int D\phi\,\left\{\phi({\bf r})\phi({\bf r}')\right\}'e^{-\Delta F_{\mathrm{dft}}[\rho,\phi]}}
{\int D\phi\,e^{-\Delta F_{\mathrm{dft}}[\rho,\phi]}}\nonumber\\
&\qquad\qquad\qquad-\overline{\phi({\bf r})\phi({\bf r}')\frac{\delta \Delta F_{\mathrm{dft}}[\rho,\phi]}{\delta\rho({\bf r})}}
\nonumber\\
&\qquad\qquad\qquad\qquad
+\overline{\phi({\bf r})\phi({\bf r}')}\left\{\overline{\frac{\delta \Delta F_{\mathrm{dft}}[\rho,\phi]}{\delta\rho({\bf r})}}\right\}\nonumber\\
&\approx
\frac{\int D\phi\,\left\{\phi({\bf r})\phi({\bf r}')\right\}'e^{-\Delta F_{\mathrm{dft}}[\rho,\phi]}}
{\int D\phi\,e^{-\Delta F_{\mathrm{dft}}[\rho,\phi]}}.
\label{dash2 app}
\end{flalign}
Accordingly, eq. (\ref{appendix phiphi transform}) reduces to
\begin{flalign}
&\overline{\frac{\delta}{\delta\rho}\iint d{\bf r}d{\bf r}'w^{-1}({\bf r}-{\bf r}')
\phi({\bf r})\phi{\bf r}')}
\nonumber\\
&\approx\int d{\bf r}'\left[
\left\{w^{-1}({\bf r}-{\bf r}')\right\}'\overline{\phi({\bf r})\phi({\bf r}')}
\right.\nonumber\\
&\hphantom{\approx\int d{\bf r}'}\qquad\qquad\qquad\qquad
\left.
+w^{-1}({\bf r}-{\bf r}')\left\{\overline{\phi({\bf r})\phi({\bf r}')}\right\}'\right]
\nonumber\\
&=\frac{1}{\rho({\bf r})}+h(0)-\int d{\bf r}'\,
h({\bf r}-{\bf r}')\rho({\bf r}')c({\bf r}'-{\bf r})-\frac{w^{-1}({\bf 0})}{\rho^2({\bf r})}\nonumber\\
&=-\int d{\bf r}'\,h({\bf r}-{\bf r}')\rho({\bf r}')c({\bf r}'-{\bf r})\nonumber\\
&=c(0)-h(0),
\label{appendix phiphi transform2}
\end{flalign}
where
\begin{flalign}
\left\{\overline{\phi({\bf r})\phi({\bf r}')}\right\}'
&=\left\{w({\bf r}-{\bf r}')\right\}'\nonumber\\
&=\left\{\frac{\delta({\bf r}-{\bf r}')}{\rho({\bf r})}-c({\bf r}-{\bf r}')\right\}'\nonumber\\
&\approx-\frac{\delta({\bf r}-{\bf r}')}{\rho^2({\bf r})},
\label{dash2}
\end{flalign}
neglecting the density dependence of the DCF, and use has been made of the Ornstein-Zernike equation in the last equality of eq. (\ref{appendix phiphi transform2}).

\section{The zero-wavevector structure factor $S(0)$ represented by the metastable DCF $c^*({\bf r})$}
\subsection{Derivation of eq. (\ref{s approx})}
We consider the Ornstein-Zernike equation for the metastable correlation functions, $c^*({\bf r})$ and $h^*({\bf r})$, in a uniform state. The Ornstein-Zernike equation at zero separation is expressed as
\begin{align}
h^*(0)&=c^*(0)
+n\int4\pi r^2dr\,c^*({\bf r})h^*({\bf r})\nonumber\\
&=c^*(0)-n\int_{r\leq\sigma}4\pi r^2dr\,c^*({\bf r})\nonumber\\
&\qquad+n\int_{r\geq\sigma}4\pi r^2dr\,c^*({\bf r})h^*({\bf r}),\label{appendix oz prep}
\end{align}
where the relation $h^*({\bf r})=-1\quad(0\leq r\leq\sigma)$ has been used in the above second line. Adding the term $-n\int_{r\geq\sigma}4\pi r^2dr\,c^*({\bf r})$ on both sides of eq. (\ref{appendix oz prep}), we have
\begin{align}
&h^*(0)-n\int_{r\geq\sigma}4\pi r^2dr\,c^*({\bf r})\nonumber\\
&=c^*(0)-n\int d{\bf r}\,c^*({\bf r})
+n\int_{r\geq\sigma}4\pi r^2dr\,c^*({\bf r})h^*({\bf r}),
\label{appendix s approx}
\end{align}
which reads
\begin{align}
&-h^*(0)-n\int d{\bf r}\,c^*({\bf r})\nonumber\\
&=-c^*(0)-n\int_{r\geq\sigma}4\pi r^2dr\,c^*({\bf r})
-n\int_{r\geq\sigma}4\pi r^2dr\,c^*({\bf r})h^*({\bf r}).
\label{appendix s approx0}
\end{align}

\subsection{Derivation of eq. (\ref{approximate structure})}
It is noted that $c^*({\bf r})h^*({\bf r})$ decays far more rapidly than $c^*({\bf r})$ even when the effective correlation functions converge to the hyperuniform ones, $c({\bf r})$ and $h({\bf r})$, for $r>\sigma$. Therefore, eq. (\ref{s approx}) becomes
\begin{align}
\frac{1}{S^*(0)}&\approx
-c^*(0)-n\int_{\sigma}^{L_c}4\pi r^2dr\,c^*({\bf r})\nonumber\\
&=-c^*(0)-\frac{6f_{\mathrm{v}}}{\pi\sigma^3}\int_{\sigma}^{L_c}
4\pi r^2dr\,c^*({\bf r})\nonumber\\
&=-c^*(0)-24f_{\mathrm{v}}\int_1^{L_c/\sigma}d\widetilde{r}\>\widetilde{r}^2c^*(\widetilde{r}),
\label{appendix sc}
\end{align}
where the relation $f_{\mathrm{v}}=\pi n\sigma^3/6$ between the volume fraction $f_{\mathrm{v}}$ and the spatially averaged density $n$ has been used in the second line of the above equation. For $\widetilde{r}=r/\sigma>1$, combination of eqs. (\ref{c result}) and (\ref{fig1 dcf}) provides
\begin{equation}
-c^*(\widetilde{r})=1-\exp\left(-\frac{\beta}{\widetilde{r}^2}\right),
\label{appendix c tilde}
\end{equation}
which will be used in calculating the last term in the last line of eq. (\ref{appendix sc}). Integration by parts yields
\begin{flalign}
&-3\int_1^{L_c/\sigma}d\widetilde{r}\>\widetilde{r}^2c^*(\widetilde{r})\nonumber\\
&=-\left[\widetilde{r}^3c^*(\widetilde{r})\right]_{1}^{L_c/\sigma}
+\int_1^{L_c/\sigma} d\widetilde{r}\,\widetilde{r}^3
\frac{dc^*(\widetilde{r})}{d\widetilde{r}}\nonumber\\
&=-\left(\frac{L_c}{\sigma}\right)^3c^*\left(\frac{L_c}{\sigma}\right)+c^*(1)
+2\beta\int_1^{L_c/\sigma}
d\widetilde{r}\,e^{-\frac{\beta}{\widetilde{r}^2}}\nonumber\\
&= I_1\left(\beta,\frac{L_c}{\sigma}\right)
+2\beta I_2\left(\beta,\frac{L_c}{\sigma}\right),
\label{appendix s by parts}
\end{flalign}
where
\begin{flalign}
I_1\left(\beta,\frac{L_c}{\sigma}\right)
=-\left(\frac{L_c}{\sigma}\right)^3c^*\left(\frac{L_c}{\sigma}\right)+c^*(1),
\label{appendix s0 i1}
\end{flalign}
and the change of variable from $\widetilde{r}$ to $x=1/\widetilde{r}$ gives
\begin{flalign}
&I_2\left(\beta,\frac{L_c}{\sigma}\right)\nonumber\\
&=\int_1^{L_c/\sigma}
d\widetilde{r}\,e^{-\frac{\beta}{\widetilde{r}^2}}\nonumber\\
&=\int^{1}_{\sigma/L_c}dx\,\frac{e^{-\beta x^2}}{x^2}\nonumber\\&=-\left[\frac{e^{-\beta
x^2}}{x}\right]^{1}_{\sigma/L_c}
-2\beta\int^{1}_{\sigma/L_c}dx\,e^{-\beta x^2}\nonumber\\
&=\frac{L_c\,e^{-\beta
(\sigma/L_c)^2}}{\sigma}-e^{-\beta}\nonumber\\
&\qquad\qquad
-\beta\sqrt{\frac{\pi}{\beta}}
\left\{\mathrm{erf}\left(\sqrt{\beta}\right)-\mathrm{erf}\left(\frac{\sqrt{\beta}\sigma}{L_c}\right)\right\}.
\label{appendix erf}
\end{flalign}
Combining eqs. (\ref{appendix s approx}) to (\ref{appendix erf}), we have
\begin{flalign}
\frac{1}{S^*(0)}=-c^*(0)+8f_{\mathrm{v}}
I_1\left(\beta,\frac{L_c}{\sigma}\right)
+16\beta f_{\mathrm{v}}
I_2\left(\beta,\frac{L_c}{\sigma}\right).
\label{appendix s0 exact}
\end{flalign}
The approximate form of eq. (\ref{appendix s0 exact}) for $L_c/\sigma\gg 1$ leads to eq. (\ref{approximate structure}):
\begin{flalign}
\frac{1}{S^*(0)}
&\approx-8f_{\mathrm{v}}\left(\frac{L}{\sigma}\right)^3c^*(L_c/\sigma)+16\beta
f_{\mathrm{v}}\frac{L_c\,e^{-\beta
(\sigma/L_c)^2}}{\sigma}\nonumber\\
&\approx24f_{\mathrm{v}}\beta\left(\frac{L_c}{\sigma}\right),
\label{appendix s0 approx}
\end{flalign}
where we have used the approximations, $-c^*(L_c/\sigma)\approx\beta(\sigma/L_c)^2$ and $L_c\,e^{-\beta (\sigma/L_c)^2}/\sigma\approx L_c/\sigma$, in the above second line. Figure 2 shows the $L_c/\sigma$--dependencies of $1/S^*(0)$ in order to compare eqs. (\ref{appendix s0 exact}) and (\ref{appendix s0 approx}).

\section{Details on the averaging operation over the $\varphi$--field in the strong-coupling approximation} 
\subsection{Shifting the fluctuating-potential field from $\phi$ to $\varphi$ in eq. (\ref{lambda av}): a general formulation and validation of eq. (\ref{conclusion varphivarphi})} 
We first see extra terms when $\Delta F_{\mathrm{dft}}[\rho,\phi]$ is represented by the $\varphi$--field. Equation (\ref{varphi def}) is rearranged to give
\begin{flalign}
\phi({\bf r})&=\varphi({\bf r})+i\Delta\psi({\bf r})-i\frac{w(0)}{2},
\label{appendix varphi def}
\end{flalign}
which further reads
\begin{flalign}
\phi({\bf r})
=\gamma\widetilde{\phi}({\bf r})
=\gamma\widetilde{\varphi}({\bf r})+i\widetilde{\Delta\psi}({\bf r})-i\frac{\gamma^2\widetilde{w}(0)}{2},
\label{appendix phi rescaling}
\end{flalign}
due to the rescaling given by eq. (\ref{chrho scaling}) and its associated form,
\begin{flalign}
\Delta\psi({\bf r})
&=-\int d{\bf r}'\gamma^2\widetilde{c}({\bf r}-{\bf r}')\frac{\widetilde{\Delta\rho}({\bf r}')}{\gamma^2}\nonumber\\
&=-\int d{\bf r}'\widetilde{c}({\bf r}-{\bf r}')\widetilde{\Delta\rho}({\bf r}')\nonumber\\
&=\widetilde{\Delta\psi}({\bf r}).
\label{appendix psi rescaling}
\end{flalign}
Plugging eq. (\ref{appendix phi rescaling}) into eq. (\ref{delta f w}), we have
\begin{flalign}
&\Delta
F_{\mathrm{dft}}\left[\rho,\phi=\gamma\widetilde{\phi}\right]\nonumber\\
&=\Delta F_{\mathrm{dft}}[\widetilde{\rho},\widetilde{\varphi}]
-\frac{1}{\gamma^2}\Delta
F_{\mathrm{dft}}[\widetilde{\rho},\widetilde{\Delta\psi}]
+\frac{i}{\gamma}\mathcal{E}_0[\widetilde{\rho},\widetilde{\varphi}]\nonumber\\
&\Delta
F_{\mathrm{dft}}\left[\widetilde{\rho},\widetilde{\Delta\psi}\right]=
\frac{1}{2}\iint d{\bf r}d{\bf r}'
\widetilde{w}^{-1}({\bf r}-{\bf r}')\widetilde{\Delta\psi}({\bf r})\widetilde{\Delta\psi}({\bf r}'),\nonumber\\
&\mathcal{E}_0[\widetilde{\rho},\widetilde{\varphi}]
=\iint d{\bf s}d{\bf s}'\,
\widetilde{w}^{-1}({\bf s}-{\bf s}')\widetilde{\varphi}({\bf s})\widetilde{\Delta\psi}({\bf s}').
\label{appendix f varphi1}
\end{flalign}
because of $\Delta F_{\mathrm{dft}}\left[\widetilde{\rho},\gamma^2\widetilde{w(0)}/2\right]=0$ for the hyperuniform correlation function $h({\bf r})$:
\begin{flalign}
&\Delta F_{\mathrm{dft}}\left[\widetilde{\rho},\frac{\gamma^2\widetilde{w(0)}}{2}\right]\nonumber\\
&=\frac{w^2(0)}{8}\int d{\bf r}\rho({\bf r})\left[
\int d{\bf r}'\left\{\delta({\bf r}-{\bf r}')+h({\bf r}-{\bf r}')\rho({\bf r}')\right\}\right]\nonumber\\
&=\frac{w^2(0)}{8}\int d{\bf r}\rho({\bf r})\left[
1+\int d{\bf r}'h({\bf r}-{\bf r}')\rho({\bf r}')\right]\nonumber\\
&=\frac{Nw^2(0)}{8}S(0)=0,
\label{appendix hyperuniform}
\end{flalign}
where $S(0)=1+\int d{\bf r}'h({\bf r}-{\bf r}')\rho({\bf r}')=0$ for the hyperuniform structure factor $S(0)$ at zero wavevector, and it is supposed that $w(0)$ is the position-independent function because of the neglect of the triplet DCF (see also the statement after eq. (\ref{s gaussian})).

Before proceeding to the $\phi$--field averaging operation defined by eq. (\ref{lambda av}), we clarify corresponding terms where the strong-coupling approximation needs to be developed. To this end, let $\zeta[\widetilde{\varphi}]$ be a general functional given by the sum of a $\widetilde{\varphi}$--independent part $\zeta_c$ and remaining $\widetilde{\varphi}$--dependent contribution $\zeta_{r}[\widetilde{\varphi}]$:
\begin{equation}
\zeta[\widetilde{\varphi}]=\zeta_c+\zeta_r[\widetilde{\varphi}].\label{appendix lambda two}
\end{equation}
It is noted that we can validate the following expansion at strong coupling (i.e., $\gamma\gg 1$):
\begin{flalign}
e^{-\frac{i}{\gamma}\mathcal{E}_0[\widetilde{\rho},\widetilde{\varphi}]}
&\approx1-\frac{1}{\gamma}\left\{i\mathcal{E}_0[\widetilde{\rho},\widetilde{\varphi}]+\frac{1}{2\gamma}\mathcal{E}_0^2[\widetilde{\rho},\widetilde{\varphi}]
\right\}\nonumber\\
&\equiv
1-\frac{1}{\gamma}\mathcal{E}[\widetilde{\rho},\widetilde{\varphi}],\nonumber\\
\mathcal{E}_0^2[\widetilde{\rho},\widetilde{\varphi}]
&\equiv\iint d{\bf r}d{\bf r}'
\widetilde{w}^{-1}({\bf r}-{\bf r}')\widetilde{\varphi}({\bf r})\widetilde{\Delta\psi}({\bf r}')\nonumber\\
&\qquad\qquad\quad\times\iint d{\bf s}d{\bf s}'\,
\widetilde{w}^{-1}({\bf s}-{\bf s}')\widetilde{\varphi}({\bf s})\widetilde{\Delta\psi}({\bf s}').
\label{appendix expansion}
\end{flalign}
Combination of eqs. (\ref{appendix f varphi1}) and (\ref{appendix expansion}) provides the approximate functional $\overline{\zeta[\widetilde{\varphi}]}$ that is averaged over the $\phi$-field, instead of the $\widetilde{\varphi}$--field, based on the definition of eq. (\ref{lambda av}) as follows:
\begin{flalign}
&\overline{\zeta[\widetilde{\varphi}]}\nonumber\\
&=\frac{\int D\phi\,\zeta[\widetilde{\varphi}]\,e^{-\Delta
F_{\mathrm{dft}}\left[\rho,\phi\right]}}
{\int D\phi\,e^{-\Delta
F_{\mathrm{dft}}\left[\rho,\phi\right]}}\nonumber\\
&=\zeta_c+\frac{\int
D\widetilde{\varphi}\,\zeta_r[\widetilde{\varphi}]\,e^{-\Delta
F_{\mathrm{dft}}\left[\rho,\phi\right]}}
{\int D\widetilde{\varphi}\,e^{-\Delta
F_{\mathrm{dft}}\left[\rho,\phi\right]}}\nonumber\\
&=\zeta_c+\frac{\int
D\widetilde{\varphi}\,\zeta_r[\widetilde{\varphi}]\,e^{-\Delta
F_{\mathrm{dft}}\left[\widetilde{\rho},\widetilde{\varphi}\right]+\frac{1}{\gamma^2}\Delta
F_{\mathrm{dft}}[\widetilde{\rho},\Delta\widetilde{\psi}]
-\frac{i}{\gamma}\mathcal{E}_0[\widetilde{\rho},\widetilde{\varphi}]}}{\int
D\widetilde{\varphi}\,e^{-\Delta
F_{\mathrm{dft}}\left[\widetilde{\rho},\widetilde{\varphi}\right]+\frac{1}{\gamma^2}\Delta
F_{\mathrm{dft}}[\widetilde{\rho},\Delta\widetilde{\psi}]
-\frac{i}{\gamma}\mathcal{E}_0[\widetilde{\rho},\widetilde{\varphi}]}}\nonumber\\
&=\zeta_c+\frac{\int
D\widetilde{\varphi}\,\zeta_r[\widetilde{\varphi}]\,\left(
1-\frac{1}{\gamma}\mathcal{E}[\widetilde{\rho},\widetilde{\varphi}]
\right)\,
e^{-\Delta
F_{\mathrm{dft}}\left[\widetilde{\rho},\widetilde{\varphi}\right]}
}{\int D\widetilde{\varphi}\,\left(
1-\frac{1}{\gamma}\mathcal{E}[\widetilde{\rho},\widetilde{\varphi}]
\right)\,e^{-\Delta
F_{\mathrm{dft}}\left[\widetilde{\rho},\widetilde{\varphi}\right]}}\nonumber\\
&=\zeta_c+\frac{\int
D\widetilde{\varphi}\,\zeta_r[\widetilde{\varphi}]\,\left(
1-\frac{1}{\gamma}\mathcal{E}[\widetilde{\rho},\widetilde{\varphi}]
\right)\,
e^{-\Delta
F_{\mathrm{dft}}\left[\widetilde{\rho},\widetilde{\varphi}\right]}
}{
\int D\widetilde{\varphi}\,e^{-\Delta
F_{\mathrm{dft}}\left[\widetilde{\rho},\widetilde{\varphi}\right]}
\left\{\frac{
\int D\widetilde{\varphi}\,\left(
1-\frac{1}{\gamma}\mathcal{E}[\widetilde{\rho},\widetilde{\varphi}]
\right)\,e^{-\Delta
F_{\mathrm{dft}}\left[\widetilde{\rho},\widetilde{\varphi}\right]}
}
{
\int D\widetilde{\varphi}\,e^{-\Delta
F_{\mathrm{dft}}\left[\widetilde{\rho},\widetilde{\varphi}\right]}
}
\right\}
}\nonumber\\
&=\zeta_c+\frac{\int
D\widetilde{\varphi}\,\zeta_r[\widetilde{\varphi}]\,\left(
1-\frac{1}{\gamma}\mathcal{E}[\widetilde{\rho},\widetilde{\varphi}]
\right)\,
e^{-\Delta
F_{\mathrm{dft}}\left[\widetilde{\rho},\widetilde{\varphi}\right]}
}{\int D\widetilde{\varphi}\,e^{-\Delta
F_{\mathrm{dft}}\left[\widetilde{\rho},\widetilde{\varphi}\right]}
\left(
1-\frac{1}{\gamma}\overline{\overline{\mathcal{E}[\widetilde{\rho},\widetilde{\varphi}]}}
\right)
}.
\label{appendix rescaled average}
\end{flalign}
Here, another averaging operation
$\overline{\overline{\mathcal{E}[\widetilde{\rho},\widetilde{\varphi}]}}$
appearing in the last line of eq. (\ref{appendix rescaled average}) has been introduced for representing
\begin{flalign}
\overline{\overline{\mathcal{E}[\widetilde{\rho},\widetilde{\varphi}]}}\equiv
\frac{\int
D\widetilde{\varphi}\,\mathcal{E}[\widetilde{\rho},\widetilde{\varphi}]\,e^{-\Delta
F_{\mathrm{dft}}\left[\widetilde{\rho},\widetilde{\varphi}\right]}}
{\int D\widetilde{\varphi}\,e^{-\Delta
F_{\mathrm{dft}}\left[\widetilde{\rho},\widetilde{\varphi}\right]}},
\label{appendix double overline def}
\end{flalign}
which becomes equivalent to eq. (\ref{lambda av}) when replacing the $\phi$--field by the $\varphi$--field; however, the notation of the above average has been altered for revealing that there is a difference between $\overline{\mathcal{E}[\widetilde{\rho},\widetilde{\varphi}]}$ and $\overline{\overline{\mathcal{E}[\widetilde{\rho},\widetilde{\varphi}]}}$ as shown by a general form (\ref{appendix rescaled average}). We obtain from eqs. (\ref{appendix expansion}) and (\ref{appendix double overline def})
\begin{flalign}
&\frac{1}{\gamma}\overline{\overline{\mathcal{E}[\widetilde{\rho},\widetilde{\varphi}]}}\nonumber\\
&=\frac{1}{\gamma}\left\{i\overline{\overline{\mathcal{E}_0[\widetilde{\rho},\widetilde{\varphi}]}}
+\frac{1}{2\gamma}\overline{\overline{\mathcal{E}_0^2[\widetilde{\rho},\widetilde{\varphi}]}}\right\}\nonumber\\
&=\frac{1}{2\gamma^2}\iiiint d{\bf r}d{\bf r}' d{\bf s}d{\bf s}'\,\overline{\overline{\widetilde{\varphi}({\bf r})\widetilde{\varphi}({\bf s})}}\nonumber\\
&\qquad\qquad\times
\widetilde{w}^{-1}({\bf r}-{\bf r}')\widetilde{\Delta\psi}({\bf r}')
\widetilde{w}^{-1}({\bf s}-{\bf s}')\widetilde{\Delta\psi}({\bf s}')\nonumber\\
&=\frac{1}{\gamma^2}\Delta
F_{\mathrm{dft}}[\widetilde{\rho},\Delta\widetilde{\psi}]
\label{appendix e average}
\end{flalign}
because of
\begin{flalign}
\overline{\overline{\widetilde{\varphi}({\bf r})}}&=0,\nonumber\\
\overline{\overline{\widetilde{\varphi}({\bf r})\widetilde{\varphi}({\bf s})}}&=\widetilde{w}({\bf r}-{\bf s}).
\label{appendix double overline varphi}
\end{flalign}
The strong-coupling approximation of the last equality in eq. (\ref{appendix rescaled average}) further validates that
\begin{flalign}
\frac{
1-\frac{1}{\gamma}\mathcal{E}[\widetilde{\rho},\widetilde{\varphi}]
}{
1-\frac{1}{\gamma}\overline{\overline{\mathcal{E}[\widetilde{\rho},\widetilde{\varphi}]}}
}
&\approx
1-\frac{1}{\gamma}\left(
\mathcal{E}[\widetilde{\rho},\widetilde{\varphi}]-\overline{\overline{\mathcal{E}[\widetilde{\rho},\widetilde{\varphi}]}}
\right)\nonumber\\
&=
1-\frac{i}{\gamma}\mathcal{E}_0[\widetilde{\rho},\widetilde{\varphi}]
-\frac{1}{2\gamma^2}\Delta\mathcal{E}_1[\widetilde{\rho},\widetilde{\varphi}]\nonumber\\
&=1-\frac{i}{\gamma}\mathcal{E}_0[\widetilde{\rho},\widetilde{\varphi}]+\mathcal{O}[\gamma^{-2}],
\label{appendix approximate division}
\end{flalign}
where
\begin{flalign}
&\Delta\mathcal{E}_1[\widetilde{\rho},\widetilde{\varphi}]
\equiv\iiiint d{\bf r}d{\bf r}' d{\bf s}d{\bf s}'\,
\left\{
\widetilde{\varphi}({\bf r})\widetilde{\varphi}({\bf s})
-w({\bf r}-{\bf s})
\right\}
\nonumber\\
&\qquad\qquad\qquad\times
\widetilde{w}^{-1}({\bf r}-{\bf r}')\widetilde{\Delta\psi}({\bf r}')
\widetilde{w}^{-1}({\bf s}-{\bf s}')\widetilde{\Delta\psi}({\bf s}').
\label{appendix approximate delta e1}
\end{flalign}
Equation (\ref{appendix approximate division}) implies that the above contribution $\Delta\mathcal{E}_1[\widetilde{\rho},\widetilde{\varphi}]$ is ignored.
Combining eqs. (\ref{appendix rescaled average}) and (\ref{appendix approximate division}), we have
\begin{flalign}
\overline{\zeta[\widetilde{\varphi}]}&=\zeta_c+\overline{\zeta_r[\widetilde{\varphi}]}\nonumber\\
&=\zeta_c+\overline{\overline{\zeta_r[\widetilde{\varphi}]}}
-\frac{i}{\gamma}\overline{\overline{\zeta_r[\widetilde{\varphi}]\mathcal{E}_0[\widetilde{\rho},\widetilde{\varphi}]}}
+\mathcal{O}[\gamma^{-2}],
\label{appendix rescaled average2}
\end{flalign}
according to the $1/\gamma$ expansion method. Incidentally, it follows from eq. (\ref{appendix rescaled average2}) that
\begin{flalign}
\overline{\widetilde{\varphi}({\bf r})\widetilde{\varphi}({\bf r}')}&=\overline{\overline{\widetilde{\varphi}({\bf r})\widetilde{\varphi}({\bf r}')}}
-\frac{i}{\gamma}\overline{\overline{\widetilde{\varphi}({\bf r})\widetilde{\varphi}({\bf r}')\mathcal{E}_0[\widetilde{\rho},\widetilde{\varphi}]}}
+\mathcal{O}[\gamma^{-2}]\nonumber\\
&=\widetilde{w}({\bf r}-{\bf r}')+\mathcal{O}[\gamma^{-2}],
\label{appendix varphivarphi}
\end{flalign}
thereby validating eq. (\ref{conclusion varphivarphi}).

In the remaining subsections, we will evaluate eqs. (\ref{ideal chem potential}) and (\ref{derivative difference sc}) based on the above strong-coupling approximation represented by eq. (\ref{appendix rescaled average2}).

 \subsection{Derivation of eq. (\ref{ideal chem potential}) from eq. (\ref{ideal diff f rhophi})} 
Following the general form (\ref{appendix lambda two}), we classify $\lambda_{\mathrm{non}}[\widetilde{\rho},\widetilde{\varphi}]$ given by eq. (\ref{ideal diff f rhophi}) into the $\widetilde{\varphi}$--independent part $\zeta_c$ and the $\widetilde{\varphi}$--dependent contribution $\gamma\zeta_1[\widetilde{\varphi}]$:
\begin{flalign}
&\lambda_{\mathrm{non}}[\widetilde{\rho},\widetilde{\varphi}]=\zeta_c+\gamma\zeta_1[\widetilde{\varphi}],\nonumber\\
&\zeta_c=\ln\frac{\widetilde{\rho}({\bf r})}{\gamma^2}+1+\frac{\gamma^2\widetilde{w}(0)}{2},\nonumber\\
&\gamma\zeta_1[\widetilde{\varphi}]=i\gamma\widetilde{\varphi}({\bf r})-\left\{
\widetilde{\rho}({\bf r})e^{\frac{\gamma^2\widetilde{w}(0)}{2}+i\gamma\widetilde{\varphi}({\bf r})}
\right\}',
\label{appendix lambda0 cr}
\end{flalign}
where the definition of $\{\cdots\}'$ is the same as that of eq. (\ref{appendix phiphi transform}).
For later convenience, we also introduce an extended form of $e^{\frac{\gamma^2\widetilde{w}(0)}{2}+i\gamma\widetilde{\varphi}({\bf r})}$:
\begin{flalign}
\gamma
u_{\mathrm{exp}}[\widetilde{\varphi}]=e^{\frac{\gamma^2\widetilde{w}(0)}{2}+i\gamma\int
d{\bf s}\,\widetilde{\varphi}({\bf s})\left\{
\widehat{\rho}_1({\bf s})+m({\bf s})
\right\}},
\label{appendix lambda exp}
\end{flalign}
where a test density field $m({\bf s})$ is added to one-particle density $\widehat{\rho}_1({\bf s})=\delta({\bf s}-{\bf r})$ that represents a single sphere located at ${\bf r}$. The functional differentiation of $u_{\mathrm{exp}}[\widetilde{\varphi}]$ with respect to $m({\bf r})$ offers the benefit of the expression (\ref{appendix lambda exp}):
\begin{flalign}
\left.
\frac{\gamma\delta u_{\mathrm{exp}}[\widetilde{\varphi}]}{\delta
m({\bf s})}\right|_{m=0}
=i\gamma\widetilde{\varphi}({\bf s})\,e^{\frac{\gamma^2\widetilde{w}(0)}{2}+i\gamma\widetilde{\varphi}({\bf r})},
\label{appendix lambda exp differentiation}
\end{flalign}
which is available to calculate the third term on the rhs of eq. (\ref{appendix rescaled average2}).

First, the Gaussian integration over the $\varphi$--field yields\begin{flalign}
\overline{\overline{\gamma\zeta_1[\widetilde{\varphi}]}}&=
-\overline{\overline{\left\{
\widetilde{\rho}({\bf r})e^{\frac{\gamma^2\widetilde{w}(0)}{2}+i\gamma\widetilde{\varphi}({\bf r})}
\right\}'}}
\nonumber\\
&=-\overline{\overline{e^{\frac{\gamma^2\widetilde{w}(0)}{2}+i\gamma\widetilde{\varphi}({\bf r})}}}
-\widetilde{\rho}({\bf r})\left\{
\frac{\overline{\overline{\delta e^{\frac{\gamma^2\widetilde{w}(0)}{2}+i\gamma\widetilde{\varphi}({\bf r})}}}}{\delta\widetilde{\rho}({\bf r})}
\right\}\nonumber\\
&=-e^{\frac{\gamma^2\widetilde{w}(0)}{2}}\overline{\overline{e^{i\gamma\int
d{\bf s}\,\widetilde{\varphi}({\bf s})\widehat{\rho}_1({\bf s})}}}
\nonumber\\
&=-e^{\frac{\gamma^2\widetilde{w}(0)}{2}-\frac{\gamma^2}{2}\iint
d{\bf s}d{\bf s}'\,\widetilde{w}({\bf s}-{\bf s}')\widehat{\rho}_1({\bf s})\widehat{\rho}_1({\bf s}')}\nonumber\\
&=-1,
\label{appendix double overline lambda r}
\end{flalign}
due to $\gamma=e^{\gamma^2\widetilde{w}(0)/2}$, $\iint d{\bf s}d{\bf s}'\,\widetilde{w}({\bf s}-{\bf s}')\widehat{\rho}_1({\bf s})\widehat{\rho}_1({\bf s}')=\widetilde{w}(0)$, $\overline{\overline{e^{\frac{\gamma^2\widetilde{w}(0)}{2}+i\gamma\widetilde{\varphi}({\bf r})}}}=1$, and the approximation similar to eq. (\ref{dash2 app}).

Next, we investigate the third term on the rhs of eq. (\ref{appendix rescaled average2}), or $\overline{\overline{\zeta_1[\widetilde{\varphi}]\mathcal{E}_0[\widetilde{\rho},\widetilde{\varphi}]}}$. The expression (\ref{appendix f varphi1}) of $\mathcal{E}_0[\widetilde{\rho},\widetilde{\varphi}]$ implies the necessity of evaluating the following contribution:
\begin{flalign}
\frac{i}{\gamma}\overline{\overline{\widetilde{\varphi}({\bf s})\gamma\zeta_1[\widetilde{\varphi}]}}
=-\overline{\overline{\widetilde{\varphi}({\bf s})\widetilde{\varphi}({\bf r})}}
-\frac{i}{\gamma}\overline{\overline{
\widetilde{\varphi}({\bf s})\,e^{\frac{\gamma^2\widetilde{w}(0)}{2}+i\gamma\widetilde{\varphi}({\bf r})}}}.
\label{appendix lambda epsilon}
\end{flalign}
It follows from eqs. (\ref{appendix double overline varphi}) and (\ref{appendix lambda exp differentiation}) that eq. (\ref{appendix lambda epsilon}) reads
\begin{flalign}
\frac{i}{\gamma}\overline{\overline{\widetilde{\varphi}({\bf s})\gamma\zeta_1[\widetilde{\varphi}]}}
=-\widetilde{w}({\bf s}-{\bf r})
-\frac{1}{\gamma}
\left.
\left(\overline{\overline{
\frac{\delta u_{\mathrm{exp}}[\widetilde{\varphi}]}{\delta
m({\bf s})}}}
\right)
\right|_{m=0}.
\label{appendix lambda epsilon2}
\end{flalign}
Since we have
\begin{flalign}
\overline{\overline{\gamma
u_{\mathrm{exp}}[\widetilde{\varphi}]}}=
e^{\frac{\gamma^2\widetilde{w}(0)}{2}-\frac{\gamma^2}{2}\iint
d{\bf s}d{\bf s}'\,\widetilde{w}({\bf s}-{\bf s}')\left\{m({\bf s})+\widehat{\rho}_1({\bf s})\right\}\left\{m({\bf s}')+\widehat{\rho}_1({\bf s}')\right\}}
\label{appendix lambda epsilon3}
\end{flalign}
in the presence of the test field $m({\bf r})$, the relation (\ref{appendix lambda exp differentiation}) reads
\begin{flalign}
&\left.
\overline{\overline{
\frac{\gamma\delta
u_{\mathrm{exp}}[\widetilde{\varphi}]}{\delta
m({\bf s})}}}
\right|_{m=0}\nonumber\\
&=-\gamma^2\left.
\int d{\bf s}'\,\widetilde{w}({\bf s}-{\bf s}')\left\{m({\bf s}')+\widehat{\rho}_1({\bf s}')\right\}\gamma
\overline{\overline{u_{\mathrm{exp}}[\widetilde{\varphi}]}}
\right|_{m=0}\nonumber\\
&=-\gamma^2\widetilde{w}({\bf s}-{\bf r})\,e^{\frac{\gamma^2\widetilde{w}(0)}{2}-\frac{\gamma^2}{2}\iint
d{\bf s}d{\bf s}'\,\widetilde{w}({\bf s}-{\bf s}')\widehat{\rho}_1({\bf s})\widehat{\rho}_1({\bf s}')}\nonumber\\&=-\gamma^2\widetilde{w}({\bf s}-{\bf r}).
\label{appendix lambda exp differentiation2}
\end{flalign}
This ensures that the first term on the rhs of eq. (\ref{appendix lambda epsilon2}) is canceled by the second term:
\begin{equation}
\overline{\overline{\widetilde{\varphi}({\bf s})\gamma\zeta_1[\widetilde{\varphi}]}}=0,
\end{equation}
thereby implying that
\begin{equation}
\overline{\overline{\zeta_1[\widetilde{\varphi}]\mathcal{E}_0[\widetilde{\rho},\widetilde{\varphi}]}}=0.
\label{appendix lambda second}
\end{equation}
Thus, we find from eqs. (\ref{appendix rescaled average2}), (\ref{appendix lambda0 cr}), (\ref{appendix double overline lambda r}) and (\ref{appendix lambda second})
\begin{flalign}
\overline{\lambda_{\mathrm{non}}[\widetilde{\rho},\widetilde{\varphi}]}
&=\zeta_c+\overline{\gamma\zeta_1[\widetilde{\varphi}]},\nonumber\\
&=\zeta_c+\overline{\overline{\gamma\zeta_1[\widetilde{\varphi}]}}
-i\overline{\overline{\zeta_1[\widetilde{\varphi}]\mathcal{E}_0[\widetilde{\rho},\widetilde{\varphi}]}}+\mathcal{O}[\gamma^{-1}]\nonumber\\
&=\zeta_c+\overline{\overline{\gamma\zeta_1[\widetilde{\varphi}]}}+\mathcal{O}[\gamma^{-1}]\nonumber\\
&\approx\ln\frac{\widetilde{\rho}({\bf r})}{\gamma^2}+1+\frac{\gamma^2\widetilde{w}(0)}{2}-1\nonumber\\
&=\ln\frac{\widetilde{\rho}({\bf r})}{\gamma^2}+\frac{\gamma^2\widetilde{w}(0)}{2}.
\label{appendix lambda0 verified}
\end{flalign}
The above last form is equivalent to the target expression
(\ref{ideal chem potential}).

\subsection{Derivation of eq. (\ref{derivative difference sc})}
Equation (\ref{delta lambda2}) reads
\begin{flalign}
\Delta\lambda[\widetilde{\rho},\widetilde{\varphi}]
&=1+i\gamma\widetilde{\varphi}({\bf r})
-\frac{\gamma\delta
U[\widetilde{\rho},\widetilde{\varphi}]/\delta\widetilde{\rho}({\bf r})}{1+U[\widetilde{\rho},\widetilde{\varphi}]/\gamma}\nonumber\\
&=1+\gamma\zeta_1[\widetilde{\varphi}]+\zeta_2[\widetilde{\varphi}]+\mathcal{O}[\gamma^{-1}],\label{appendix
delta lambda}
\end{flalign}
where $\zeta_1[\widetilde{\varphi}]$ has been given in eq. (\ref{appendix lambda0 cr}) and
\begin{flalign}
\zeta_2[\widetilde{\varphi}]
&=-\int d{\bf r}_2\widetilde{\rho}({\bf r}_2)e^{-v({\bf r}_1-{\bf r}_2)}
\left\{
\widetilde{\rho}({\bf r}_1)e^{i\gamma\int d{\bf s}\,\widetilde{\varphi}({\bf s})\widehat{\rho}_2({\bf s})}
\right\}'\nonumber\\
&+\int d{\bf r}'\widetilde{\rho}({\bf r}')\,e^{i\gamma\int d{\bf s}\,\widetilde{\varphi}({\bf s})\widehat{\rho'}_1({\bf s})}
\left\{
\widetilde{\rho}({\bf r})
e^{i\gamma\int d{\bf s}\,\widetilde{\varphi}({\bf s})\widehat{\rho}_1({\bf s})}
\right\}',
\end{flalign}
where the definition of $\{\cdots\}'$ is the same as that of eq. (\ref{appendix phiphi transform}).
The non-equilibrium chemical potential difference $\Delta\lambda[\rho]$ is obtained from averaging $\overline{\Delta\lambda[\widetilde{\rho},\widetilde{\varphi}]}$ over the $\widetilde{\varphi}$--field in a similar manner to the strong-coupling approximation adopted in eq. (\ref{appendix lambda0 verified}):
\begin{flalign}
&\Delta\lambda[\rho]=\overline{\Delta\lambda[\widetilde{\rho},\widetilde{\varphi}]}\nonumber\\
&=
1+\overline{\overline{\gamma\zeta_1[\widetilde{\varphi}]}}+
\overline{\overline{\zeta_2[\widetilde{\varphi}]}}
-i\overline{\overline{\zeta_1[\widetilde{\varphi}]
\mathcal{E}_0[\widetilde{\rho},\widetilde{\varphi}]}}+\mathcal{O}[\gamma^{-1}]\nonumber\\
&=
1-1+
\overline{\overline{\zeta_2[\widetilde{\varphi}]}}+\mathcal{O}[\gamma^{-1}]
\nonumber\\
&\approx-\int d{\bf r}_2\,\widetilde{\rho}({\bf r}_2)e^{-v({\bf r}_1-{\bf r}_2)}
\overline{\overline{
e^{i\gamma\int d{\bf s}\,\widetilde{\varphi}({\bf s})\widehat{\rho}_2({\bf s})}
}}\nonumber\\
&\quad-\int d{\bf r}_2\,\widetilde{\rho}({\bf r}_2)e^{-v({\bf r}_1-{\bf r}_2)}\widetilde{\rho}({\bf r}_1)
\frac{\delta
\overline{\overline{
e^{i\gamma\int d{\bf s}\,\widetilde{\varphi}({\bf s})\widehat{\rho}_2({\bf s})}
}}
}{
\delta\widetilde{\rho}({\bf r}_1)
}\nonumber\\
&\qquad\qquad+\int d{\bf r}'\widetilde{\rho}({\bf r}')\,
\overline{\overline{
e^{i\gamma\int d{\bf s}\,\widetilde{\varphi}({\bf s})\left\{\widehat{\rho}_1({\bf s})+\widehat{\rho'}_1({\bf s})\right\}}
}},
\label{appendix delta lambda av}
\end{flalign}
where we have used the results, $\overline{\overline{\gamma\zeta_1[\widetilde{\varphi}]}}=-1$ and $\overline{\overline{\zeta_1[\widetilde{\varphi}]\mathcal{E}_0[\widetilde{\rho},\widetilde{\varphi}]}}=0$, given by eqs. (\ref{appendix double overline lambda r}) and (\ref{appendix lambda second}), respectively, and $\delta\{\overline{\overline{e^{i\gamma\int d{\bf s}\,\widetilde{\varphi}\left\{\widehat{\rho}_1+\widehat{\rho'}_1\right\}}}}\}/\delta\rho=0$ due to the neglect of the density dependence of $e^{-w(0)}=1/\gamma^2$ as before. It follows from the Gaussian integration when performing the averages in the last equality of eq. (\ref{appendix delta lambda av}) that
\begin{flalign}
&\overline{\overline{
e^{i\gamma\int d{\bf s}\,\widetilde{\varphi}({\bf s})\widehat{\rho}_2({\bf s})}
}}\nonumber\\
&=e^{-\frac{\gamma^2}{2}\iint d{\bf s}d{\bf s}'\,\widetilde{w}({\bf s}-{\bf s}')\widehat{\rho}_2({\bf s})\widehat{\rho}_2({\bf s}')}\nonumber\\
&=e^{-\gamma^2\iint d{\bf s}d{\bf s}'\,\widetilde{w}({\bf s}-{\bf s}')\left\{
\delta({\bf s}-{\bf r}_1)\delta({\bf s}'-{\bf r}_2)+\sum_{i=1}^2\frac{\delta({\bf s}-{\bf r}_i)\delta({\bf s}'-{\bf r}_i)}{2}
\right\}}\nonumber\\
&=e^{-\gamma^2\widetilde{w}({\bf r}_1-{\bf r}_2)-\gamma^2\widetilde{w}(0)}\nonumber\\
&=\frac{1}{\gamma^2}e^{-\gamma^2\widetilde{w}({\bf r}_1-{\bf r}_2)},
\label{appendix two particle av}\\
&\frac{\delta\overline{\overline{
e^{i\gamma\int d{\bf s}\,\widetilde{\varphi}({\bf s})\widehat{\rho}_2({\bf s})}
}}
}{\delta\widetilde{\rho}({\bf r}_1)}
=\frac{\delta({\bf r}_1-{\bf r}_2)}{\widetilde{\rho}^2({\bf r}_1)}e^{-\gamma^2\widetilde{w}({\bf r}_1-{\bf r}_2)},
\label{appendix derivative av}
\end{flalign}
and
\begin{flalign}
&\overline{\overline{
e^{i\gamma\int d{\bf s}\,\widetilde{\varphi}({\bf s})\left\{\widehat{\rho}_1({\bf s})+\widehat{\rho'}_1({\bf s})\right\}}}}
\nonumber\\
&=e^{-\frac{\gamma^2}{2}\iint d{\bf s}d{\bf s}'\,\widetilde{w}({\bf s}-{\bf s}')\left\{
\widehat{\rho}_1({\bf s})\widehat{\rho}_1({\bf s}')+\widehat{\rho'}_1({\bf s})\widehat{\rho'}_1({\bf s}')
\right\}}\nonumber\\
&=e^{-\frac{\gamma^2\widetilde{w}(0)}{2}-\frac{\gamma^2\widetilde{w}(0)}{2}}\nonumber\\
&=\frac{1}{\gamma^2},
\label{appendix one particle av}
\end{flalign}
respectively. In eq. (\ref{appendix one particle av}), it is noted that the cross terms vanish: $\widehat{\rho}_1({\bf s})\widehat{\rho'}_1({\bf s}')=\widehat{\rho'}_1({\bf s})\widehat{\rho}_1({\bf s}')=0$ because the one-particle densities, $\widehat{\rho}_1({\bf s})=\delta({\bf s}-{\bf r})$ and $\widehat{\rho'}_1({\bf s})=\delta({\bf s}-{\bf r}')$, represent instantaneous densities of a target particle located at different positions of ${\bf r}$ and ${\bf r}'$ and a single sphere is unable to simultaneously exist at different locations.

Substituting eqs. (\ref{appendix two particle av}) to (\ref{appendix one particle av}) into eq. (\ref{appendix delta lambda av}), we obtain
\begin{flalign}
&\Delta\lambda[\rho]=-\frac{1}{\gamma^2}\int d{\bf r}_2\,\widetilde{\rho}({\bf r}_2)e^{-v({\bf r}_1-{\bf r}_2)-w({\bf r}_1-{\bf r}_2)}\nonumber\\
&\hphantom{\Delta\lambda[\rho]=-\frac{1}{\gamma^2}\int d{\bf r}_2}
-e^{-v(0)-\gamma^2\widetilde{w}(0)}+\frac{1}{\gamma^2}\int d{\bf r}'\widetilde{\rho}({\bf r}')\nonumber\\
&=\frac{1}{\gamma^2}\int d{\bf r}'\left\{1-e^{-v({\bf r}-{\bf r}')-\gamma^2\widetilde{w}({\bf r}-{\bf r}')}\right\}\widetilde{\rho}({\bf r}')-\frac{e^{-v(0)}}{\gamma^2}\nonumber\\
&=\int d{\bf r}'\left\{1-e^{-v({\bf r}-{\bf r}')-w({\bf r}-{\bf r}')}\right\}\rho({\bf r}')+\mathcal{O}[\gamma^{-2}],
\label{appendix delta lambda av result}
\end{flalign}
hence verifying eq. (\ref{derivative difference sc}).

\section*{Conflicts of interest}
There are no conflicts to declare.

\section*{Acknowledgements}
The author thanks the anonymous referees for their valuable comments and suggestions.

\bibliographystyle{apsrev4-1}

\begin{thebibliography}{}
\bibitem{t revjcp} S. Torquato, J. Chem. Phys., 2018, {\bf 149}, 020901. 
\bibitem{t physrep} S. Torquato, Phys. Rep., 2018, {\bf 745}, 1. 
\bibitem{t mat} J. Kim and S. Torquato, P. Natl. Acad. Sci. USA, 2020, {\bf 117}, 8764-8774. 
\bibitem{h mat2020} G. J. Aubry, L. S. Froufe-P\'erez, U. Kuhl, O. Legrand, F. Scheffold and F. Mortessagne, Phys. Rev. Lett., 2020, {\bf 125}, 127402. 

\bibitem{t first} S. Torquato and F. H. Stillinger, Phys. Rev. E, 2003, {\bf 68}, 041113. 
\bibitem{t donev1} A. Donev, F. H. Stillinger and S. Torquato, Phys. Rev. Lett., 2005, {\bf 95}, 090604. 
\bibitem{t donev2} A. Donev, S. Torquato and F. H. Stillinger, Phys. Rev. E, 2005, {\bf 71}, 011105. 
\bibitem{t disk} S. Atkinson, F. H. Stillinger and S. Torquato, P. Natl. Acad. Sci. USA, 2014, {\bf 52}, 18436-18441. 
\bibitem{t void1} C. E. Zachary, Y. Jiao and S. Torquato, Phys. Rev. Lett., 2011, {\bf 106}, 178001.
\bibitem{t void2} C. E. Zachary, Y. Jiao and S. Torquato, Phys. Rev. E, 2011, {\bf 83}, 051308. 
\bibitem{t rattler} S. Atkinson, F. H. Stillinger and S. Torquato, Phys. Rev. E, 2013, {\bf 88}, 062208. 
\bibitem{t marginal} Y. Kallus and S. Torquato, Phys. Rev. E, 2014, {\bf 90}, 022114.
\bibitem{t dcf} S. Atkinson, F. H. Stillinger and S. Torquato, Phys. Rev. E, 2016, {\bf 94}, 032902. 
\bibitem{t slowing} S. Atkinson, G. Zhang, A. B. Hopkins and S. Torquato, Phys. Rev. E, 2016, {\bf 94}, 012902. 
\bibitem{t kim} J. Kim and S. Torquato, Phys. Rev. B, 2018, {\bf 97}, 054105. 
\bibitem{t real} G. Zhang and S. Torquato, Phys. Rev. E, 2020, {\bf 101}, 032124.
\bibitem{t preprint} S. Torquato, Phys. Rev. E, 2021, {\bf 103}, 052126.

\bibitem{hecke rev} M. van Hecke, J. Phys.: Condens. Matt., 2009, {\bf 22}, 033101. 
\bibitem{liu rev} A. J. Liu and S. R. Nagel, Annu. Rev. Condens. Matt. Phys., 2010, {\bf 1}, 347-369. 
\bibitem{wyart rev} M. M\"uller and M. Wyart, M., Annu. Rev. Condens. Matt. Phys., 2015, {\bf 6}, 177-200. 
\bibitem{lub2017}
V. Lubchenko and P. G. Wolynes, J. Phys. Chem. B, 2017, {\bf 122}, 3280-3295. 
\bibitem{lub2018} V. Lubchenko, Adv. Phys. X, 2018, {\bf 3}, 1510296. 

\bibitem{ikeda2017} H. Mizuno, H. Shiba and A. Ikeda, P. Natl. Acad. Sci. USA, 2017, {\bf 114}, E9767-E9774. 
\bibitem{ikeda2018} M. Shimada, H. Mizuno, M. Wyart and A. Ikeda, Phys. Rev. E, 2018, {\bf 98}, 060901. 
\bibitem{ikeda2020 prr} H. Mizuno, M. Shimada and A. Ikeda, Phys. Rev. Res., 2020, {\bf 2}, 013215. 
\bibitem{ikeda2020 jcp} H. Mizuno, H. Tong, A. Ikeda and S. Mossa, J. Chem. Phys., 2020, {\bf 153}, 154501. 
\bibitem{ikeda2020 sm} M. Shimada, H. Mizuno and A. Ikeda, Soft Matter, 2020, {\bf 16}, 7279-7288. 
\bibitem{lerner2013} E. Lerner, G. D\"uring and M. Wyart, Soft Matter, 2013, {\bf 9}, 8252-8263. 
\bibitem{lerner2016} E. Lerner, G. D\"uring and E. Bouchbinder, Phys. Rev. Lett., 2016, {\bf 117}, 035501. 
\bibitem{lerner2020 prl} D. Richard, K. Gonz\'alez-L\'opez, G. Kapteijns, R. Pater, T. Vaknin, E. Bouchbinder and E. Lerner, Phys. Rev. Lett., 2020, {\bf 125}, 085502. 
\bibitem{lerner2020 pre} A. Moriel, Y. Lubomirsky, E. Lerner and E. Bouchbinder, Phys. Rev. E, 2020, {\bf 102}, 033008. 
\bibitem{lerner2020 pnas} C. Rainone, E. Bouchbinder and E. Lerner, P. Natl. Acad. Sci. USA, 2020, {\bf 117}, 5228-5234. 
\bibitem{urbani2021} C. Rainone, P. Urbani, F. Zamponi, E. Lerner and E. Bouchbinder, SciPost Phys., 2021, {\bf 4}, 008. 
\bibitem{berthier2016} L. Berthier, P. Charbonneau, Y. Jin, G. Parisi, B. Seoane and F. Zamponi, P. Natl. Acad. Sci. USA, 2016, {\bf 113}, 8397-8401. 
\bibitem{szamel nc} L. Wang, A. Ninarello, P. Guan, L. Berthier, G. Szamel and E. Flenner, Nat. Commun., 2019, {\bf 10}, 1-7. 
\bibitem{berthier2019} L. Berthier, G. Biroli, P. Charbonneau, E. I. Corwin, S. Franz and F. Zamponi, J. Chem. Phys., 2019, {\bf 151}, 010901. 
\bibitem{tan sm2021} X. Tan, Y. Guo, D. Huang and L. Zhang, Soft Matter, 2021, {\bf 17}, 1330-1336. 
\bibitem{ozawa} M. Ozawa, L. Berthier and D. Coslovich, SciPost Phys., 2017, {\bf 3}, 027. 
\bibitem{chieco} A. T. Chieco, M. Zu, A. J. Liu, N. Xu and D. J. Durian, Phys. Rev. E, 2018, {\bf 98}, 042606. 
\bibitem{silbert} L. E. Silbert and M. Silbert, Phys. Rev. E, 2009, {\bf 80}, 041304. 
\bibitem{olsson} Y. Wu, P. Olsson and S. Teitel, Phys. Rev. E, 2015, {\bf 92}, 052206. 
\bibitem{ikeda non1} A. Ikeda and L. Berthier, Phys. Rev. E, 2015, {\bf 92}, 012309.
\bibitem{ikeda non2} A. Ikeda, L. Berthier and G. Parisi, Phys. Rev. E, 2017, {\bf 95}, 052125. 

\bibitem{witt} M. te Vrugt, H. L\"owen and R. Wittkowski, Adv. Phys., 2020, {\bf 69}, 121-247. 
\bibitem{dean} D. S. Dean, J. Phys. A: Math. Gen., 1996, {\bf 29}, L613. 
\bibitem{seifert} T. Leonard, B. Lander, U. Seifert and T. Speck, J. Chem. Phys., 2013, {\bf 139}, 204109. 
\bibitem{kim} B. Kim, K. Kawasaki, H. Jacquin and F. van Wijland, Phys. Rev. E, 2014, {\bf 89}, 012150. 
\bibitem{jac} H. Jacquin, B. Kim, K. Kawasaki and F. van Wijland, Phys. Rev. E, 2015, {\bf 91}, 022130. 
\bibitem{das2015} N. Bidhoodi and S. P. Das, Phys. Rev. E, 2015, {\bf 92}, 012325.
\bibitem{podgornik} D. S. Dean, B. S. Lu, A. C. Maggs and R. Podgornik, Phys. Rev. Lett., 2016, {\bf 116}, 240602. 
\bibitem{demery} V. D\'emery and D. S. Dean, J. Stat. Mech.: Theo. Exp., 2016, {\bf 2016}, 023106. 
\bibitem{kruger} M. Kr\"uger and D. S. Dean, J. Chem. Phys., 2017, {\bf 146}, 134507. 
\bibitem{lut science} J. F. Lutsko, Sci. Adv., 2019, {\bf 5}, eaav7399. 
\bibitem{frusawa1} H. Frusawa, J. Phys. A: Math. Theo., 2019, {\bf 52}, 065003. 
\bibitem{frusawa2} H. Frusawa, Entropy, 2020, {\bf 22}, 34. 
\bibitem{goles} S. Mahdisoltani and R. Golestanian, 2021, Phys. Rev. Lett., {\bf 126}, 158002. 

\bibitem{ry} T. V. Ramakrishnan and M. Yussouff, Phys. Rev. B, 1979, {\bf 19}, 2775. 
\bibitem{evans} R. Evans, Adv. Phys., 1979, {\bf 28}, 143-200. 
\bibitem{singh} Y. Singh, Phys. Rep., 1991, {\bf 207}, 351-444. 
\bibitem{likos} C. N. Likos, Phys. Rep., 2001, {\bf 348}, 267-439. 
\bibitem{lut1} J. F. Lutsko, Adv. Chem. Phys., 2010, {\bf 144}, 1. 
\bibitem{singh1985} Y. Singh, J. P. Stoessel and P. G. Wolynes, Phys. Rev. Lett., 1985, {\bf 54}, 1059. 
\bibitem{baus} M. Baus and J. L. Colot, J. Phys. C: Solid State Phys., 1986, {\bf 19}, L135. 
\bibitem{woly1987} R. W. Hall and P. G. Wolynes, J. Chem. Phys., 1987, {\bf 86}, 2943-2948. 
\bibitem{dasgupta1992} C. Dasgupta, Europhys. Lett., 1992, {\bf 20}, 131. 
\bibitem{bagchi} R. K. Murarka and B. Bagchi, J. Chem. Phys., 2001, {\bf 115}, 5513-5520. 
\bibitem{das2001} C. Kaur and S. P. Das, Phys. Rev. Lett., 2001, {\bf 86}, 2062. 
\bibitem{munakata} K. Kim and T. Munakata, Phys. Rev. E, 2003, {\bf 68}, 021502. 
\bibitem{sood} P. Chaudhuri, S. Karmakar, C. Dasgupta, H. R. Krishnamurthy and A. K. Sood, Phys. Rev. Lett., 2005, {\bf 95}, 248301. 
\bibitem{dasgupta2008} P. Chaudhuri, S. Karmakar and C. Dasgupta, Phys. Rev. Lett., 2005, {\bf 100}, 125701. 
\bibitem{das2012} B. S. Gupta, L. Premkumar and S. P. Das, Phys. Rev. E, 2012, {\bf 85}, 051501. 
\bibitem{das2016} L. Premkumar, N. Bidhoodi and S. P. Das, J. Chem. Phys., 2001, {\bf 144}, 124511. 
\bibitem{odagaki} T. Odagaki, J. Phys. Soc. Jpn., 2017, {\bf 86}, 082001. 
\bibitem{das2020} A. Mondal and S. P. Das, Prog. Theor. Exp. Phys., 2020, {\bf 2020}, 073I02. 

\bibitem{behringer rev} R. P. Behringer and B. Chakraborty, Rep. Prog. Phys., 2018, {\bf 82}, 012601. 
\bibitem{makse}
A. Baule, F. Morone, H. J. Herrmann and H. A. Makse, Rev. Mod. Phys., 2018, {\bf 90}, 015006. 

\bibitem{b char2015} P. Charbonneau, E. I. Corwin, G. Parisi and F. Zamponi, Phys. Rev. Lett., 2015, {\bf 114}, 125504. 
\bibitem{b franz2015} S. Franz, G. Parisi, P. Urbani and F. Zamponi, P. Natl. Acad. Sci. USA, 2015, {\bf 112}, 14539-14544. 
\bibitem{b char2016} P. Charbonneau, E. I. Corwin, G. Parisi, A. Poncet and F. Zamponi, Phys. Rev. Lett., 2016, {\bf 117}, 045503. 
\bibitem{b hexner} D. Hexner, A. J. Liu and S. R. Nagel, Phys. Rev. E, 2018, {\bf 97}, 063001. 
\bibitem{b franz2020} S. Franz, A. Sclocchi and P. Urbani, SciPost Phys., 2020, {\bf 9}, 012. 
\bibitem{b char2020} P. Charbonneau, E. I. Corwin, R. C. Dennis, R. D. H. Rojas, H. Ikeda, G. Parisi and F. Ricci-Tersenghi, Phys. Rev. E, 2021, {\bf 104}, 014102.
\bibitem{b liu} S. A. Ridout, J. W. Rocks and A. J. Liu, arXiv preprint, 2020, arXiv:2011.13049. 

\bibitem{zamponi2012} P. Charbonneau, E. I. Corwin, G. Parisi and F. Zamponi, Phys. Rev. Lett., 2012, {\bf 109}, 205501. 
\bibitem{zamponi2014} P. Charbonneau, J. Kurchan, G. Parisi, P. Urbani and F. Zamponi, 2014, Nat. Commun., 2014, {\bf 5}, 1-6. 
\bibitem{berthier prx} Q. Liao and L. Berthier, Phys. Rev. X, 2019, {\bf 9}, 011049. 
\bibitem{parisi2020} C. Artiaco, P. Baldan and G. Parisi, Phys. Rev. E, 2020, {\bf 101}, 052605. 
\bibitem{still jcp} F. H. Stillinger, P. G. Debenedetti and S. Sastry, J. Chem. Phys., 1998, {\bf 109}, 3983-3988. ,
\bibitem{still nat} P. G. Debenedetti and F. H. Stillinger, Nature, 2001, {\bf 410}, 259-267. 
\bibitem{heuer} A. Heuer, Journal of Physics: Condens. Matt., 2008, {\bf 20}, 373101. 
\bibitem{ediger rev} M. D. Ediger and P. Harrowell, J. Chem. Phys., 2012, {\bf 137}, 080901. 

\bibitem{frusawa3} H. Frusawa,  J. Stat. Mech.: Theo. Exp., 2021, {\bf 2021}, 013213.

\bibitem{py} D. Henderson, Condens. Matt. Phys., 2009, {\bf 12}, 127.

\bibitem{frydel}
D. Frydel and M. Ma, Phys. Rev. E, 2016, {\bf 93}, 062112.

\bibitem{frusawa4}
H. Frusawa, Phys. Rev. E, 2020, {\bf 101}, 012121.

\bibitem{netz} R. R. Netz, Eur. Phys. J. E, 2001, {\bf 5}, 557. 

\bibitem{s gab} A. Gabrielli, Phys. Rev. E, 2004, {\bf 70}, 066131. 
\bibitem{s chak prl} J. N. Nampoothiri, Y. , Wang, K. Ramola, J. Zhang, S. Bhattacharjee and B Chakraborty, Phys. Rev. Lett., 2020, {\bf 125}, 118002. 
\bibitem{s tanaka} H. Tong, S. Sengupta and H. Tanaka, Nat. Commun., 2020, {\bf 11}, 1-10. 
\bibitem{lerner2020  jcp} E. Lerner, J. Chem. Phys., 2020, {\bf 153}, 216101. 
\bibitem{shimada stress} M. Shimada and E. De Giuli, arXiv preprint, 2020, arXiv:2008.11896. 

\bibitem{two div}D. Hexner, A. J. Liu and S. R. Nagel, Phys. Rev. Lett., 2018, {\bf 121}, 115501.
\bibitem{parisi2021}P. Rissone, E. I. Corwin and G Parisi, Phys. Rev. Lett., 2021, {\bf 127}, 038001.


\bibitem{frusawa0} H. Frusawa and R. Hayakawa, J. Phys. A: Math. Gen., 2000, {\bf 33}, L155. 

\end{thebibliography}
 \end{document}